\documentclass[showpacs,twocolumn,superscriptaddress]{revtex4-1}
\usepackage{doi}
\usepackage{hyperref}
\hypersetup{
  colorlinks=true,        
  linkcolor=blue,         
  citecolor=cyan,         
}

\usepackage{graphicx}
\usepackage{dcolumn}
\usepackage{bm}
\usepackage{color}
\usepackage{enumitem}
\usepackage{amsmath}
\usepackage{amssymb}
\usepackage{orcidlink}
\usepackage{subcaption}
\usepackage{graphicx} 
\begin{document}
\title{
Regular black hole's impact on the gravitational waveforms from periodic orbits\\
}

\author{Mirzabek Alloqulov}
\email{malloqulov@gmail.com}
\affiliation{School of Physics, Harbin Institute of Technology, Harbin 150001, People’s Republic of China}

\author{Sanjar Shaymatov}
\email{sanjar@astrin.uz}
\affiliation{Institute of Fundamental and Applied Research, National Research University TIIAME, Kori Niyoziy 39, Tashkent 100000, Uzbekistan}
\affiliation{Institute for Theoretical Physics and Cosmology,
Zhejiang University of Technology, Hangzhou 310023, China}
\affiliation{University of Tashkent for Applied Sciences, Str. Gavhar 1, Tashkent 100149, Uzbekistan}

\author{Bobomurat Ahmedov}
\email{ahmedov@astrin.uz}
\affiliation{School of Physics, Harbin Institute of Technology, Harbin 150001, People’s Republic of China}
\affiliation{Institute for Advanced Studies, New Uzbekistan University, Movarounnahr str. 1, Tashkent 100000, Uzbekistan}
\affiliation{Institute of Theoretical Physics, National University of Uzbekistan, Tashkent 100174, Uzbekistan}

\author{Tao Zhu}
\email{zhut05@zjut.edu.cn}
\affiliation{Institute for Theoretical Physics and Cosmology, Zhejiang University of Technology, Hangzhou 310023, China}
\affiliation{United Center for Gravitational Wave Physics (UCGWP), Zhejiang University of Technology, Hangzhou, 310032, China}
\date{\today}

\begin{abstract}
In this paper, we investigate periodic orbits exhibiting zoom-whirl behavior around a magnetically charged black hole (MCBH) within the framework of the regular black hole. We examine how the magnetic charge influences orbital dynamics by modifying the background spacetime geometry, thereby affecting the energy and angular momentum of particles. In particular, we calculate the radii of the marginally bound orbits (MBOs) and innermost stable circular orbits (ISCOs), demonstrating that the magnetic charge parameter reduces both radii. This provides valuable insight into the role of the charge parameter in shaping orbital behavior and altering spacetime geometry. We model the complex motion of a stellar-mass object as a timelike particle, inspiraling into a supermassive black hole (SMBH) in the MCBH background, with its trajectory described using periodic geodesic orbits. Based on this analysis of such periodic orbits, we further analyze the gravitational waveforms generated by extreme mass ratio inspirals (EMRIs), in which the SMBH's spacetime dominates the dynamics of the stellar-mass object. By combining particle trajectory analysis with waveform modeling in a semi-analytical approach, we show that the charge parameter significantly alters the zoom-whirl orbital dynamics and induces notable changes in the waveform structure. These results illustrate that future gravitational wave (GW) observations may constrain the properties of MCBHs, thereby deepening our understanding of the gravitational imprint of regular black holes.
\end{abstract}

\maketitle

\section{Introduction}

Black holes (BHs), exact solutions to Einstein's field equations and fascinating predictions of general relativity (GR) made long ago, have undergone a revolutionary transformation. From theoretical speculation, they have become observational realities nearly a century later \cite{Abbott16a, Abbott16b, Akiyama19L1, Akiyama22L12}. This shift is driven by modern observational advancements and data analysis, such as the first detection of gravitational waves (GWs) from merging BHs \cite{Abbott16a, Abbott16b} and the first image of a BH's shadow, the supermassive BH at the center of galaxy M87 and Sgr A$^\star$ \cite{Akiyama19L1, Akiyama22L12}. 
These modern observations confirmed GR's predictions and the existence of BHs, allowing us to test GR's limits and gain a deeper understanding of gravity \cite{Will14LRR, Psaltis20PRL}.

While remarkably successful, GR encounters a fundamental limitation: it cannot resolve the singularities predicted within BHs, where the theory breaks down. This inability to explain curvature singularities has been a long-standing difficulty of GR \cite{Hawking1970}. Furthermore, GR's limitations in explaining phenomena like dark matter and the accelerated expansion of the universe confirm its incompleteness. Motivated by these issues, promising alternative theories have emerged, such as those that introduce regular BHs \cite{Bardeen68, Eyon, Hayward06}.  
These approaches are crucial for finding fundamental solutions to singularities and achieving a more complete description of spacetime near BHs \cite{Ayon-Beato2000PLB, Bronnikov2000}. Nonlinear electrodynamics, among these alternatives, features nonlinear interactions between electromagnetic fields. These interactions can alter the spacetime geometry in the vicinity of BH centers, resulting in the formation of regular BHs that are characterized by a smooth, de Sitter-like core rather than singularities (see, e.g., \cite{Bardeen68B, Dymnikova92}). Importantly, this approach maintains consistency, together with key energy conditions \cite{Balart14PLB, Kumar20MNRAS}. Based on these foundational studies, significant progress has been made in developing BH solutions that eliminate central singularities \cite{Fan16, Toshmatov17, Neves14PLB, Narzilloev20b, Ghosh21AP}. These advances address key theoretical challenges and offer predictions for novel observational signatures \cite{Shaymatov23ApJ, Rayimbaev22IJMPD, Xamidov24EPJC}, particularly through analyses of geodesic and horizon structures \cite{Perlick18}. More recently, this line of research has sparked increased research activity, with various studies exploring the influence of NED parameters in various astrophysical scenarios (see, e.g., \cite{Bisnovatyi-Kogan17, Rogers15MNRAS, Atamurotov21JCAP, Kumar20MNRAS, Jafarzade25APh, Shaymatov22PRD, Xamidov25JCAP, Si-Jiang25JCAP}).

Through the first detection of GWs \cite{Abbott16a, Abbott16b}, GW astronomy has not only verified the existence of BHs but also opened new possibilities to examine the validity of Einstein’s GR and explore alternative theories of gravity such as the curvature of spacetime \cite{1916SPAW.......189S}. After this direct detection of GWs, numerous such phenomena have been recorded through ground-based GW detectors within the frequency range of $10-10^3$ Hz in strong gravitational field regimes \cite{Abbott19aPRX, Abbott24:PhysRevD.109.022001, Abbott23PRX}. It is also worth emphasizing that space-based GW observatories like LISA \cite{Amaro-Seoane2017LISA}, Taiji \cite{Hu:10.1093/nsr/nwx116}, and TianQin \cite{Luo:TianQin2016}, which are currently under development, will allow detecting GW events within the mHz frequency band \cite{Baibhav21ExA, Amaro-Seoane2023LRR, Arun22LRR:LISA}, thereby opening new possibilities in GW astronomy and enabling tests of potential deviations from GR in this frequency band. In addition, their unique sensitivity in this low-frequency range will enable the exploration of new astrophysical sources and provide crucial tools for testing deviations in GW physics. Therefore, space-based GW observatories are particularly well-suited to detecting signals from stellar-mass compact objects orbiting supermassive black holes (SMBHs), usually referred to as extreme mass ratio inspirals (EMRIs) \cite{Hughes2001CQG, Amaro-Seoane18LRR, Babak17PRD}. These systems emit GWs in the low-frequency band, making them the main targets for detectors like LISA, Taiji, and Tianqin. EMRIs, which involve a stellar-mass object gradually inspiraling into an SMBH, provide unique opportunities to test the nature of BHs in various gravity models in the strong-field regime \cite{Amaro-Seoane2007CQG, McGee2020NatAs, Laghi2021MNRAS, Amaro-Seoane2022GRG}. Moreover, the gravitational radiation from EMRIs may carry rich information about the dynamics near BH event horizons and can shed light on the properties of regular BHs, serving as a powerful tool for probing both the fundamental nature of BHs and their surrounding environments. 

The gravitational waveforms emitted by EMRIs are modeled by analyzing the complex motion of a stellar-mass object orbiting an SMBH, where the extreme mass ratio causes the background spacetime of the SMBH to dominate the smaller object’s dynamics. The trajectory of a smaller object can often be described using periodic orbits. In this context, the properties of gravitational waveforms can be understood by examining the periodic geodesic orbits of the inspiraling object around Schwarzschild and Kerr BHs (see, e.g. \cite{Levin_2008, Grossman_2009, Misra_2010}). These bound trajectories often exhibit zoom-whirl behavior, in which the compact object completes an integer number of radial and angular oscillations during each orbit. These periodic orbits include important features such as the zoom-whirl behavior, where the object undergoes multiple rapid revolutions near the BH before moving outward again. Therefore, such orbits are categorized by three integers, such as zoom ($z$), whirl ($w$), and vertex number ($v$), and are further defined by the ratio of their angular to radial frequencies \cite{Levin_2008, Levin_2009}. These quantities offer a powerful framework for classifying the orbital dynamics and interpreting the gravitational waveforms from EMRIs \cite{Glampedakis02PRD}. On these lines, an extensive analysis has been carried out in Refs.~\cite{Levin2010, Babar17PRD, Liu_2019, Deng20, Wei2019PRD, Jiang2024PDU, Tu23PRD, Chen:2025aqh, Lu:2025cxx, meng2025periodicorbitsgravitationalwave}, examining how the nature of periodic orbits depends on the BH background geometry in various gravity models. Similarly, regular BHs can also influence orbital dynamics, which in turn alters the resulting EMRI gravitational waveforms \cite{Barausse14PRD, Cardoso22PRD, Yang2025JCAP, Haroon2025, Alloqulov25GW1}.

This paper presents a detailed analysis of periodic geodesic orbits around a magnetically charged regular black hole (MCBH) \cite{Ali_2020}. We explore the key characteristics of such orbits and the corresponding gravitational waveforms produced by EMRIs. Using waveform modeling for inspiraling stellar-mass objects \cite{Babak07PRD}, we develop a theoretical framework intended for comparison with observational data from future space-based detectors such as LISA, Taiji, and TianQin \cite{Amaro-Seoane2017LISA, Hu:10.1093/nsr/nwx116, Luo:TianQin2016}.

The paper is organized as follows. In Section~\ref{Sec:II}, we begin with a brief review of the MCBH, followed by an analysis of how the magnetic charge parameter affects orbital dynamics and the radii of ISCOs and MBOs for particles with appropriate parameters. In Section~\ref{Sec:III}, we study periodic orbits around the MCBH and examine the features of such periodic orbits under the influence of the magnetic charge parameter. Finally, we study the gravitational waveforms emitted from the EMRI system through the periodic orbits around the MCBH in Section~\ref{Sec:IV}. We end up with a conclusion and a discussion in Section~\ref{Sec:V}.

\section{Spacetime and particle dynamics}\label{Sec:II}
Here, we briefly review the magnetically charged black hole (MCBH) spacetime and examine the geodesic motion of particles in this background~\cite{Ali_2020}. The spacetime metric can be written in the following form: 
\begin{equation}
    ds^2=-f(r)dt^2+\frac{dr^2}{f(r)}+r^2(d\theta^2+\sin^2{\theta}d\phi^2)\ ,
\end{equation}
with
\begin{equation}
    f(r)=1-\frac{2M}{r} e^{\frac{-q^2}{2Mr}}\ ,
\end{equation}
where $q$ is the magnetic charge of the BH. If $q$ tends to zero, we can recover the Schwarzschild spacetime. 
\begin{figure}[htbp]
    \centering
    \includegraphics[scale=0.5]{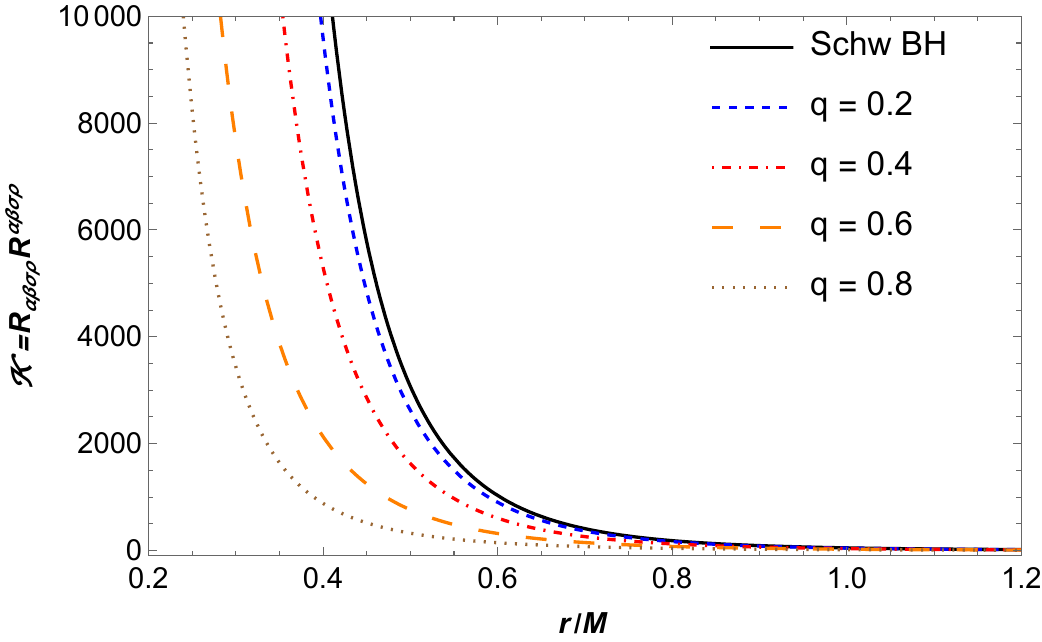}
    \caption{The plot illustrates the radial dependence of the Kretschmann scalar for different values of the magnetic charge parameter.}
    \label{fig:kr}
\end{figure}
\begin{figure*}[htbp]
    \includegraphics[scale=0.5]{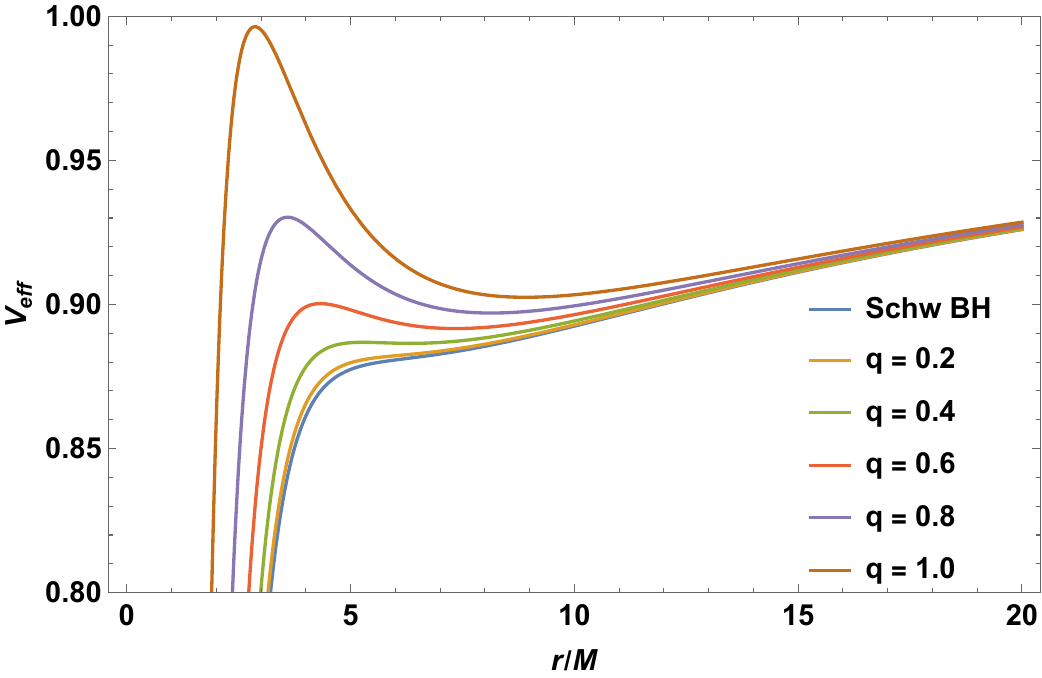}
    \includegraphics[scale=0.5]{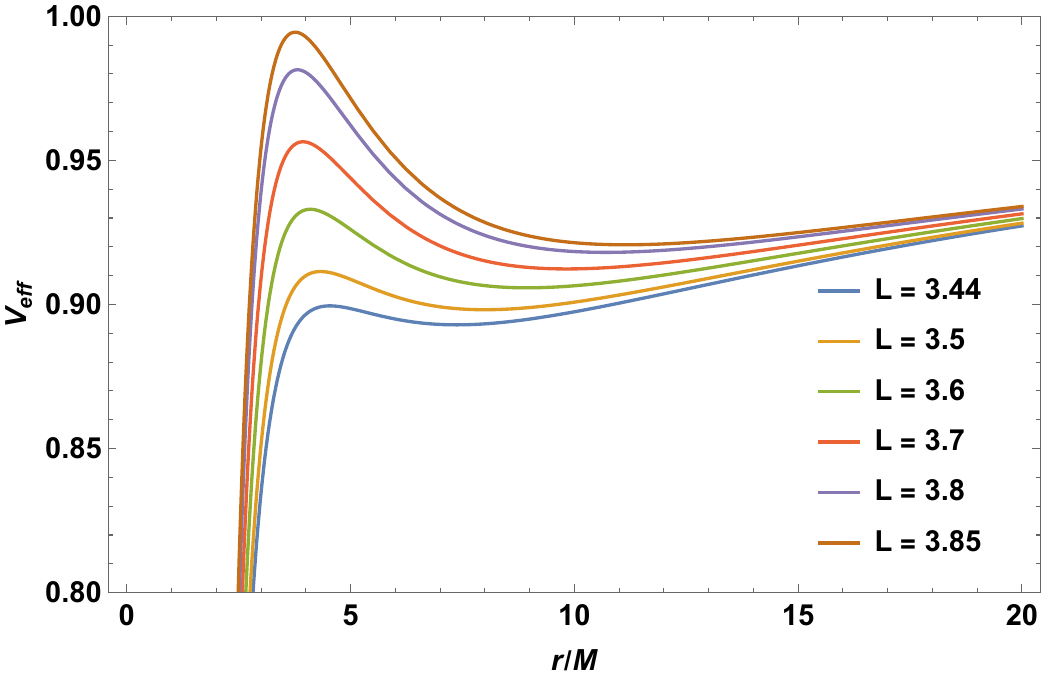}
    \caption{The plot shows the radial dependence of the effective potential for different values of the magnetic charge parameter $q$ (left panel) and the orbital angular momentum (right panel).}
    \label{fig:eff}
\end{figure*}
\begin{figure}[htbp]
    \centering
    \includegraphics[scale=0.5]{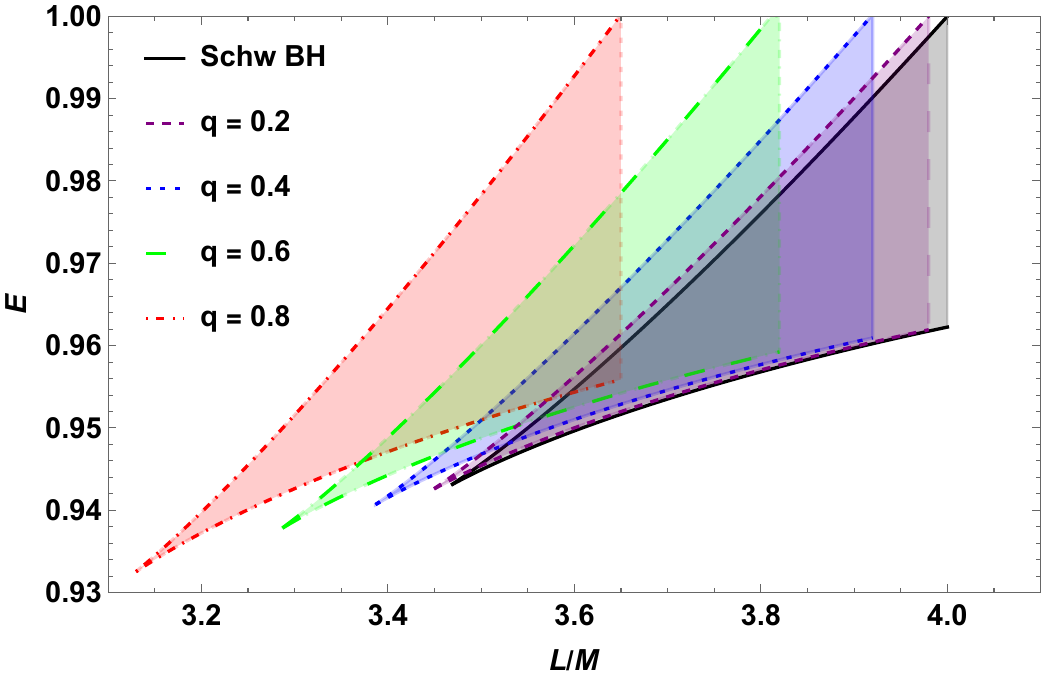}
    \caption{The plot shows the allowed parameter space of the energy and orbital angular momentum for the bound orbits around the MCBH with different values of the magnetic charge parameter $q$}.
    \label{fig:EvsL}
\end{figure}

One can calculate the Kretschmann scalar to analyze the spacetime singularity as follows: 
\begin{equation}
\mathcal{K}=R_{\mu\nu\sigma\rho}R^{\mu\nu\sigma\rho}\ , 
\end{equation}
which takes the following form for the MCBH
\begin{eqnarray}
\mathcal{K}&=&\frac{e^{-\frac{q^2}{M r}}}{4 M^2 r^{10}}\Big(192 M^4 r^4-192 M^3 r^3 q^2+ \\ \nonumber
&+&96 M^2 r^2 q^4 - 16 M r q^6+q^8\Big)\, . 
\end{eqnarray}
It should be emphasized that that we can recover the result for the Schwarzschild BH when considering $q=0$. Then it reads as follows:
\begin{eqnarray}\label{krch2}
\lim_{q \to 0}\mathcal{K}&=& \frac{48 M^2}{r^6}\, .
\end{eqnarray}
As demonstrated by Eq.~(\ref{krch2}), the real physical singularity exists at $r=0$. To be more informative, we demonstrate the radial dependence of the Kretschmann scalar for different values of the magnetic charge parameter in Fig.~\ref{fig:kr}.
Furthermore, we explore the massive particle dynamics around the MCBH. The Lagrangian of the test particle can be written in the following form~\cite{1983mtbh.book.....C}
\begin{equation}
\mathcal{L}=\frac{1}{2}m\,g_{\mu\nu}\,\frac{dx^{\mu}}{d\tau}\,\frac{dx^{\nu}}{d\tau},
\end{equation}
where $m$ and $\tau$ are the mass of the test particle and the proper time, respectively. It is worth noting that we set the mass of the test particle as $m=1$ for simplicity. After that, we can write the generalized momentum of the particle as~\cite{1983mtbh.book.....C}:
\begin{eqnarray}
p_{\mu}=\frac{\partial {\cal L}}{\partial \dot{x}^{\mu}}=g_{\mu \nu}\dot{x}^{\nu}\, .
\end{eqnarray}
The equations of motion can be written as follows, using the above equation
\begin{eqnarray}\label{eq:eqmotion}
p_{t} &=& -\left(1-\frac{2M}{r} e^{\frac{-q^2}{2Mr}}\right)\dot{t}=-E \, , \nonumber\\
p_{\phi} &=& r^{2}\sin^{2}\theta\dot{\phi}=L \, , \nonumber \\
p_{r} &=& \left(1-\frac{2M}{r} e^{\frac{-q^2}{2Mr}}\right)^{-1}\dot{r} \, , \nonumber\\
p_{\theta} &=& r^{2}\dot{\theta} \, ,
\end{eqnarray}
where $L$ and $E$ refer to the orbital angular momentum and energy of the particle, respectively. 
It is worth noting that the motion in the equatorial plane is considered ($\theta=\pi/2$). Using the normalization condition, one can write the following expression 
\begin{equation}
\frac{\dot{r}^{2}}{1-\frac{2M}{r} e^{\frac{-q^2}{2Mr}}}+\frac{L^{2}}{r^{2}}-\frac{E^{2}}{1-\frac{2M}{r} e^{\frac{-q^2}{2Mr}}}=-1.
\end{equation}
It can be rewritten in a simpler form as follows
\begin{equation}
\dot{r}^{2}+V_{\rm eff}=E^{2},
\end{equation}
with
\begin{equation}
V_{\rm eff}=\left(1-\frac{2M}{r} e^{\frac{-q^2}{2Mr}}\right)\left(1+\frac{L^{2}}{r^{2}}\right)\, ,
\end{equation}
where $V_{eff}$ refers to the effective potential for the motion of the massive particle. In Fig.~\ref{fig:eff}, the radial dependence of the effective potential is demonstrated. One can see from this figure that the values of the effective potential increase under the influence of the magnetic charge parameter $q$ and the orbital angular momentum. 

\begin{table*}[]
\centering
\resizebox{0.55\textwidth}{!}{
\begin{tabular}{cccccc}
\hline
\hline
$q$ & $L_{MBO}$ & $r_{MBO}$ & $L_{ISCO}$ & $r_{ISCO}$ & $E_{ISCO}$  \\ \hline
0.0 & 4.00000 & 4.00000 & 3.46410 & 6.00000 & 0.942809 \\ \hline
 0.2 & 3.97990 & 3.95985 & 3.44476 & 5.93979 & 0.942279 \\ \hline
 0.4 & 3.91833 & 3.83748 & 3.38551 & 5.75645 & 0.940603 \\ \hline
 0.6 & 3.81100 & 3.62630 & 3.28224 & 5.44078 & 0.937482 \\ \hline
 0.8 & 3.64864 & 3.31156 & 3.12606 & 4.97232 & 0.932215 \\ \hline \hline
\end{tabular}
}
\caption{The values of the MBO and ISCO parameters are tabulated for different values of the magnetic charge parameter.}
\label{table2}
\end{table*}
\renewcommand{\arraystretch}{1.2}
\begin{table*}[]
\centering
\resizebox{1.0\textwidth}{!}{
\begin{tabular}{cccccccccc}
\hline
\hline
$q$ & $L_{av}$ & $E_{(1,1,0)}$ & $E_{(1,2,0)}$ & $E_{(2,1,1)}$ & $E_{(2,2,1)}$ & $E_{(3,1,2)}$ & $E_{(3,2,2)}$ & $E_{(4,1,3)}$ & $E_{(4,2,3)}$ \\ \hline
0.0    & 3.73205 &  0.965425  &  0.968383  &  0.968026  &  0.968434  &  0.968225  &  0.968438  & 0.968285   & 0.968440   \\ \hline
0.2    & 3.71233 & 0.965096   &  0.968099  & 0.967736   &  0.968151  & 0.967938   &  0.968156  &  0.967999  & 0.968157   \\ \hline
0.4    & 3.65192 &  0.964049  & 0.967201   &  0.966816  &  0.967258  & 0.96703   &  0.967263  & 0.967095   & 0.967264    \\ \hline
0.6    & 3.54662 &  0.962081  & 0.965533   & 0.965101   & 0.965598   & 0.96534   &  0.965603  & 0.965413   &   0.965605 \\ \hline
0.8    & 3.38735 & 0.958688   & 0.962722   & 0.962198   & 0.962805   &  0.962485  &  0.962812  &  0.962574 &    0.962814   \\ \hline \hline
\end{tabular}
}
\caption{The values of the energy $E$ are tabulated for different periodic orbits characterized by $(z,w,v)$. The magnetic charge parameter varies from $q=0$ to $q=0.8$ with an interval of $0.2$. Here, we set $L_{av}=\frac{1}{2}(L_{MBO}+L_{ISCO})$.}
\label{table1}
\end{table*}
In this paper, we mainly focus on the periodic orbits, which are a type of bound orbits, around the MCBH. The following conditions can be applied for energy and orbital angular momentum when a particle moves along a generally bounded orbit
\begin{equation}
    L_{ISCO}\leq L \quad\mbox{and}\quad E_{ISCO}\leq E \leq E_{MBO}=1\, ,
\end{equation}
with $L_{ISCO}$ and $E_{ISCO}$, which are the orbital angular momentum and the energy of the particle that is moving on ISCO, respectively. $E_{MBO}$ refers to the energy of the particle moving on MBO. We need to explore the MBO and ISCO to investigate the effect of the magnetic charge parameter on the properties of the periodic orbits. One can determine the MBO using the following conditions
\begin{equation}
    V_{eff}=1\quad\mbox{and}\quad \frac{d V_{eff}}{dr}=0.
\end{equation}
We numerically calculate the values of the MBO parameters for different values of the magnetic charge of the MCBH, and the results are represented in Table~\ref{table2}. As can be demonstrated from this table that the MBO parameters decrease with the increase of the magnetic charge parameter $q$. Then we turn to the ISCO and determine the ISCO parameters using the following conditions
\begin{equation}
    \dot{r}=0, \quad \frac{d V_{eff}}{dr}=0 \quad\mbox{and}\quad \frac{d^2V_{eff}}{dr^2}=0.
\end{equation}
The values of the ISCO parameters, which are the energy, the orbital angular momentum, and the radius, are calculated for different values of the magnetic charge parameter in Table~\ref{table2}. One can analyze from this table that the values of the ISCO parameters decrease as the magnetic charge parameter $q$ increases. It is worth noting that $q=0$ refers to the Schwarzschild BH case. To be more informative, we demonstrate the allowed parameter space of the energy and orbital angular momentum for bound orbits around the MCBH for different values of the magnetic charge parameter in Fig.~\ref{fig:EvsL}. It can be seen from this figure that the region shifts to the left as the increase of magnetic charge parameter, meaning that bound orbits around the MCBH with large values of the magnetic charge parameter have a lower orbital angular momentum boundary when the energy is fixed.

\section{Periodic orbits around the magnetically charged black hole}\label{Sec:III}
\begin{figure*}
    \includegraphics[scale=0.5]{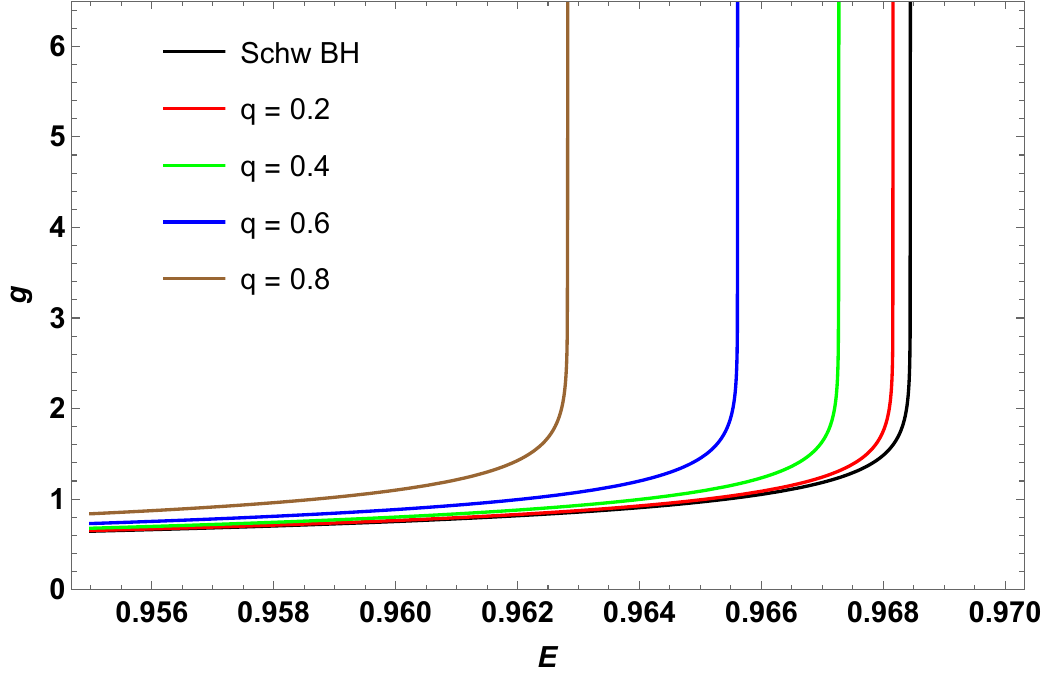}
    \includegraphics[scale=0.49]{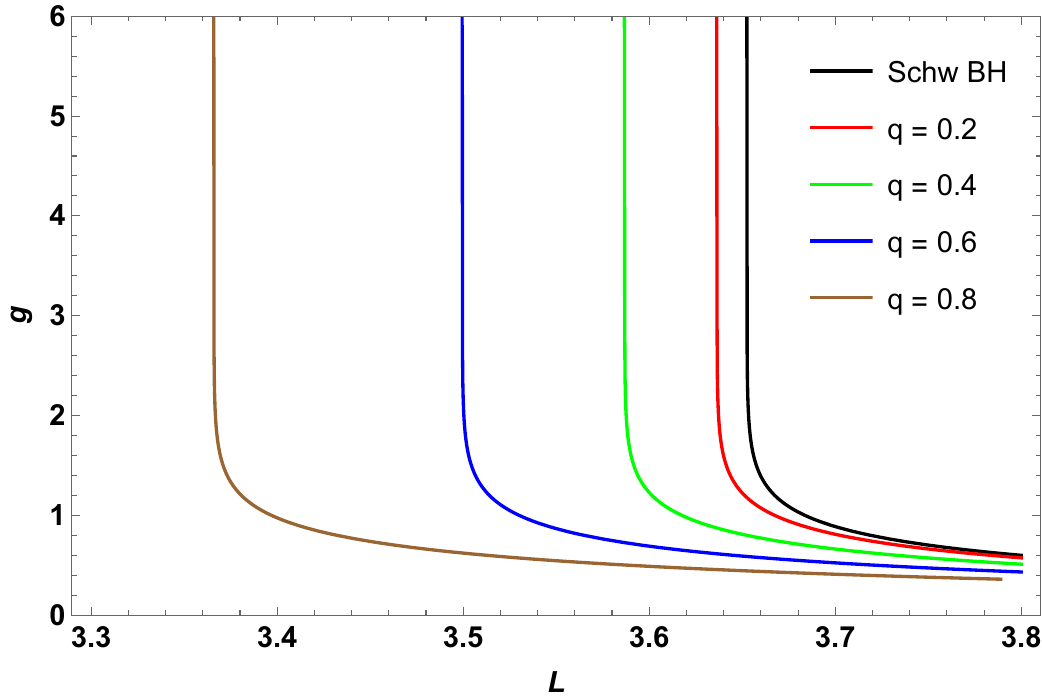}
    \caption{Left panel: The plot shows the rational number as a function of the energy of the particles for different values of the magnetic charge parameter. Here, we set $L_{av}=\frac{1}{2}(L_{MBO}+L_{ISCO})$. Right panel: The plot illustrates the rational number as a function of the orbital angular momentum for different values of the magnetic charge parameter $q$. The energy of the particles is $E=0.96$ for this panel. }
    \label{g}
\end{figure*}
\begin{figure*}[htbp]
    \centering
    \includegraphics[width=0.32\textwidth]{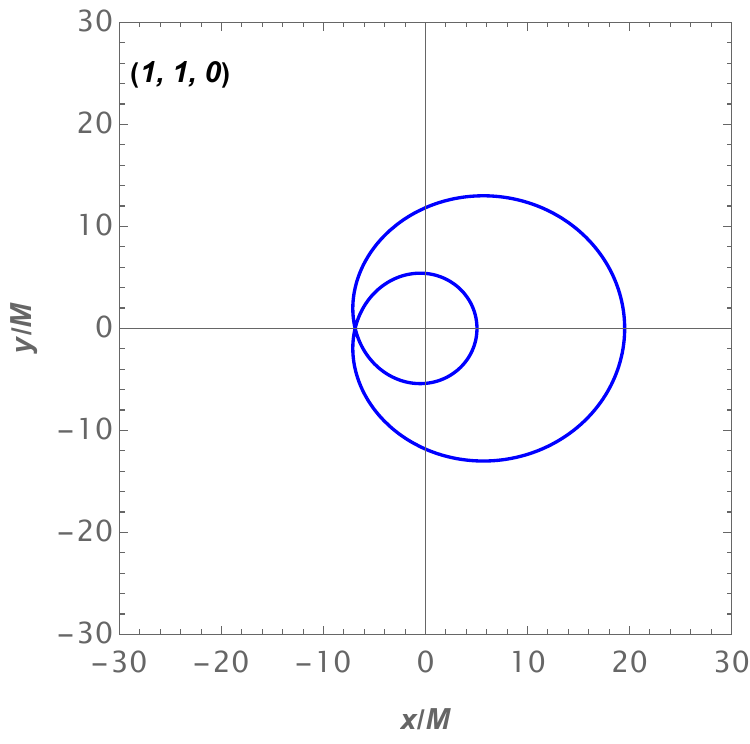} \hfill
    \includegraphics[width=0.32\textwidth]{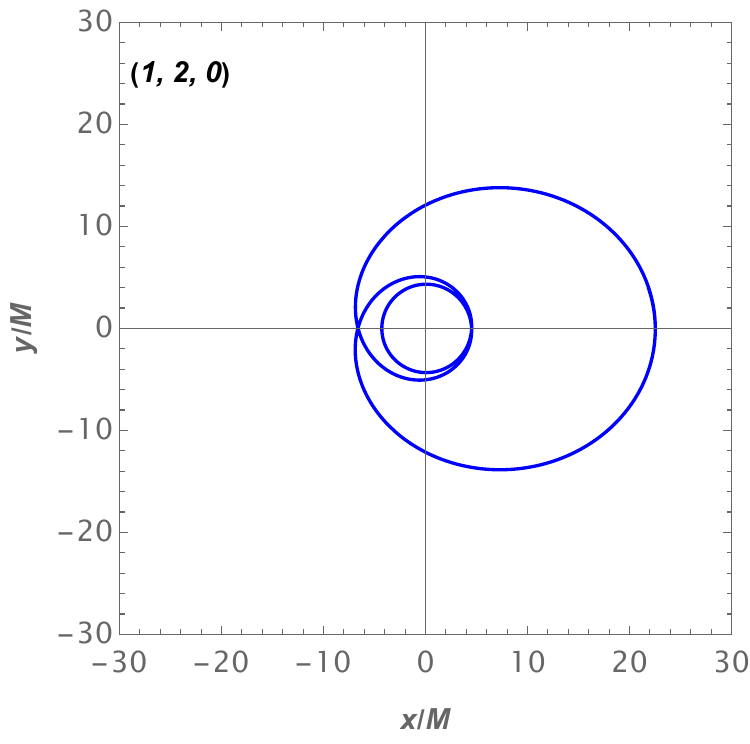} \hfill
    \includegraphics[width=0.32\textwidth]{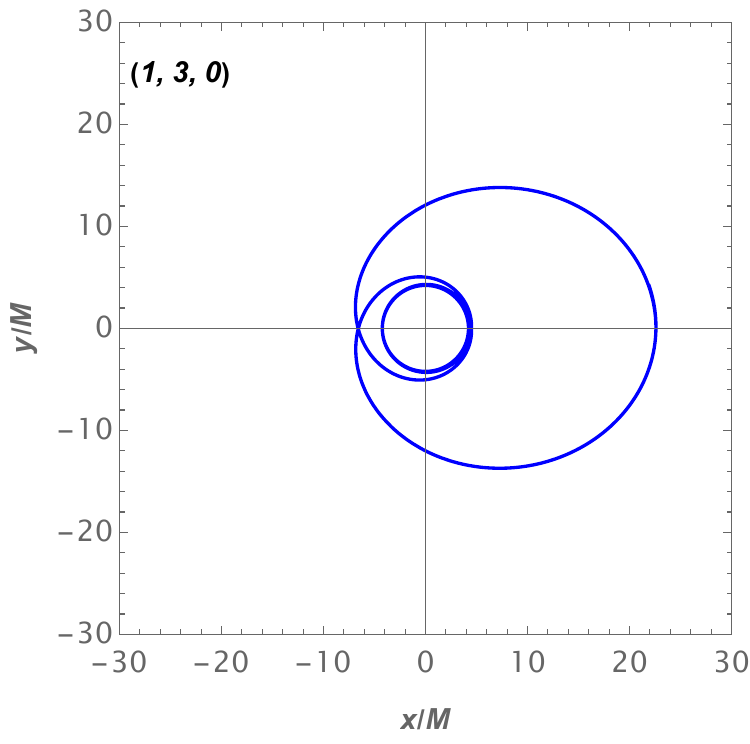} \\
    
    \vspace{0.2cm} 
    \includegraphics[width=0.32\textwidth]{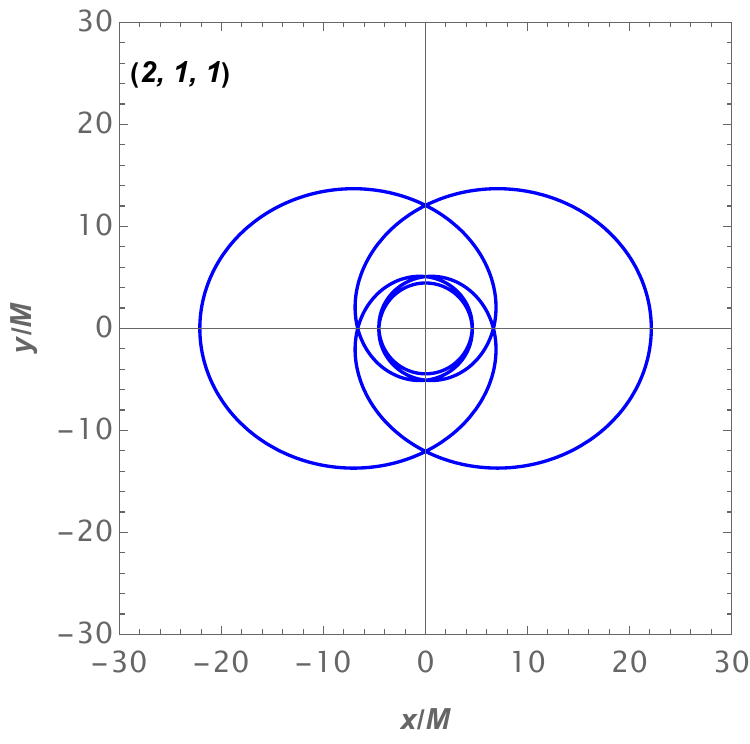} \hfill
    \includegraphics[width=0.32\textwidth]{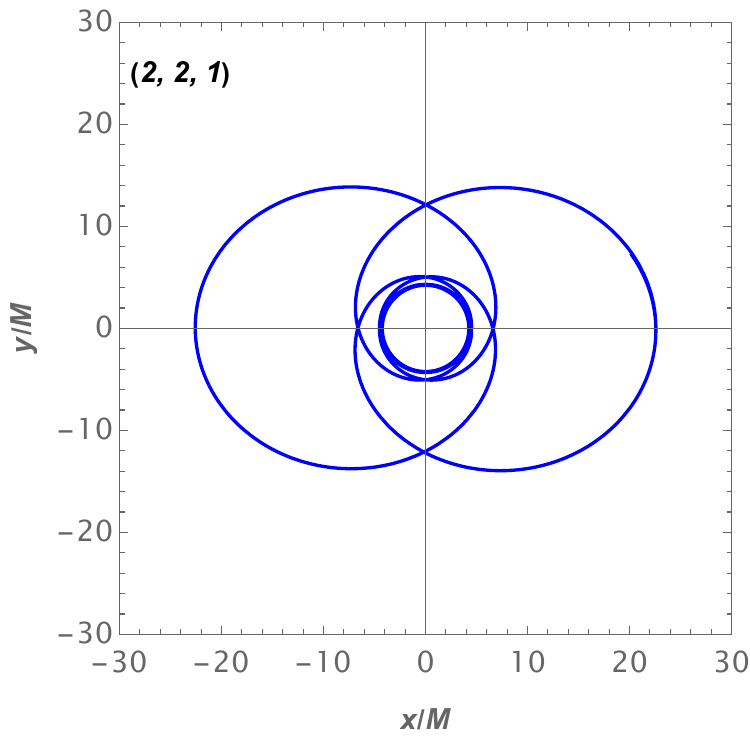} \hfill
    \includegraphics[width=0.32\textwidth]{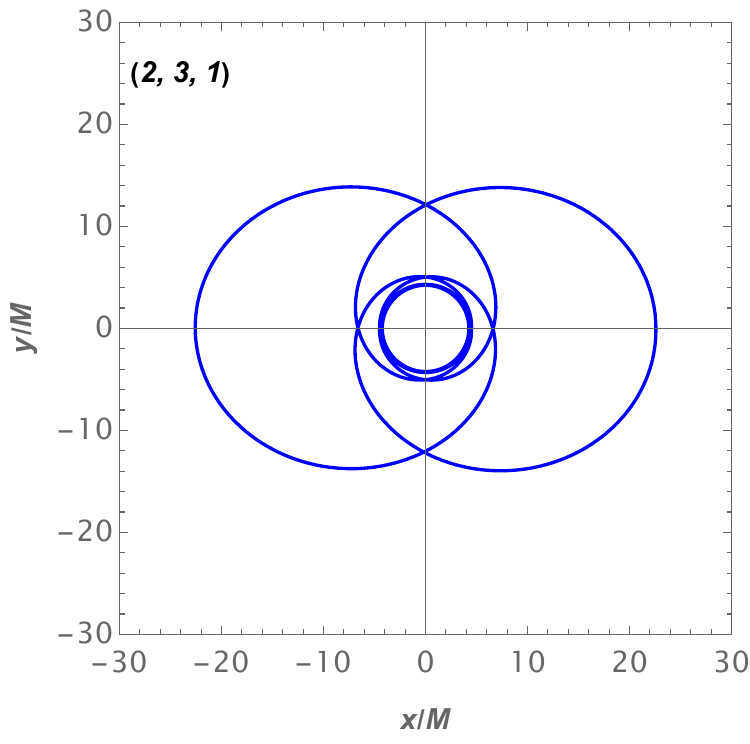} \\
    
    \vspace{0.2cm} 
    \includegraphics[width=0.32\textwidth]{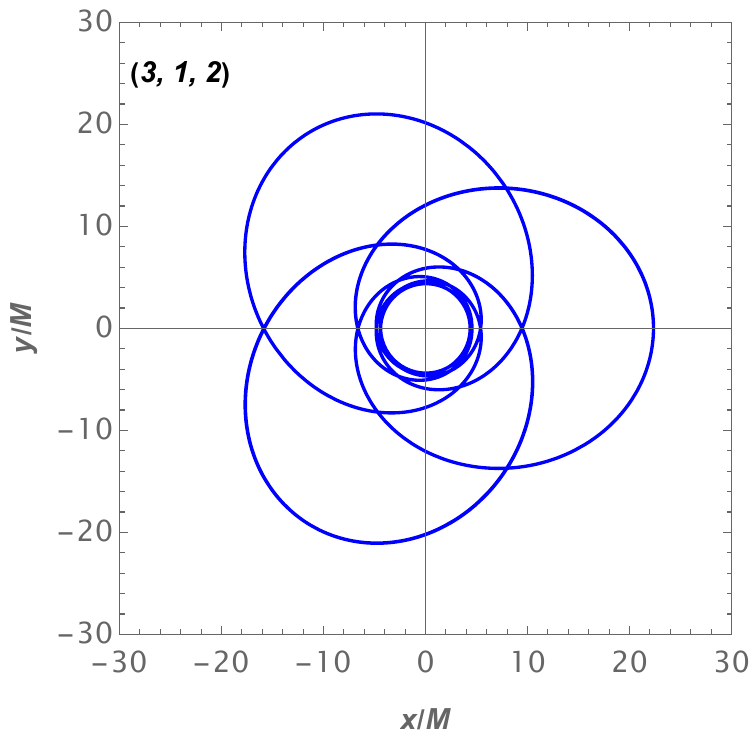} \hfill
    \includegraphics[width=0.32\textwidth]{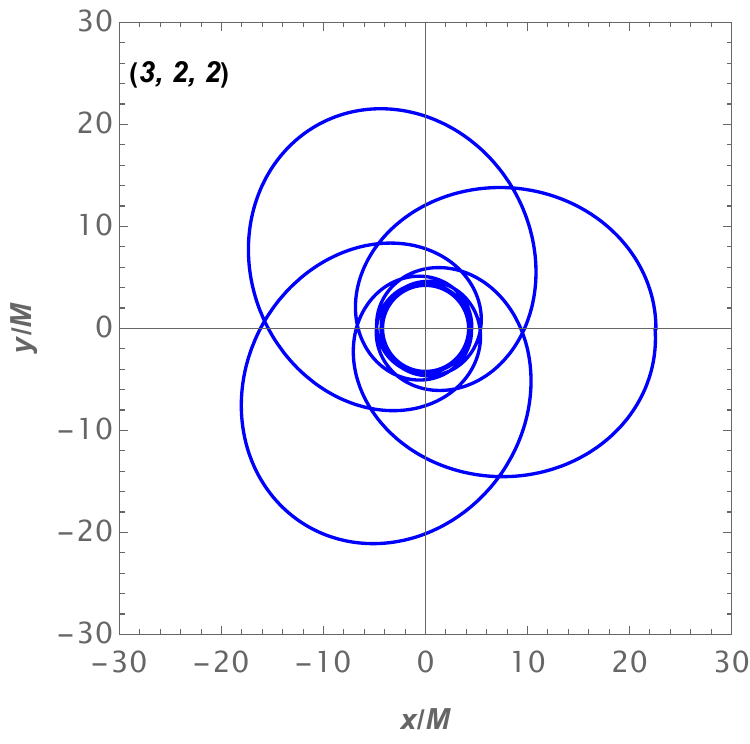} \hfill
    \includegraphics[width=0.32\textwidth]{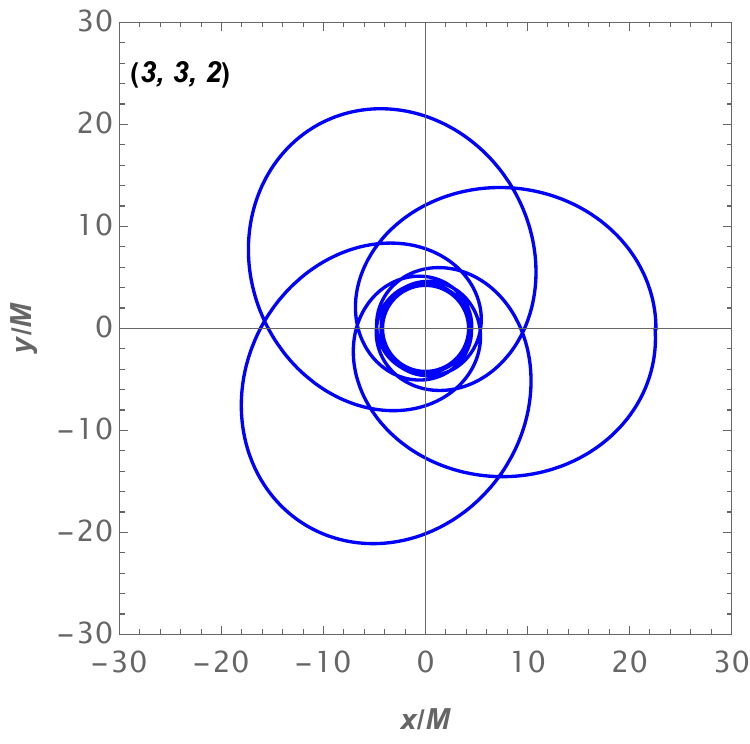} \\
    
    \vspace{0.2cm} 
    \includegraphics[width=0.32\textwidth]{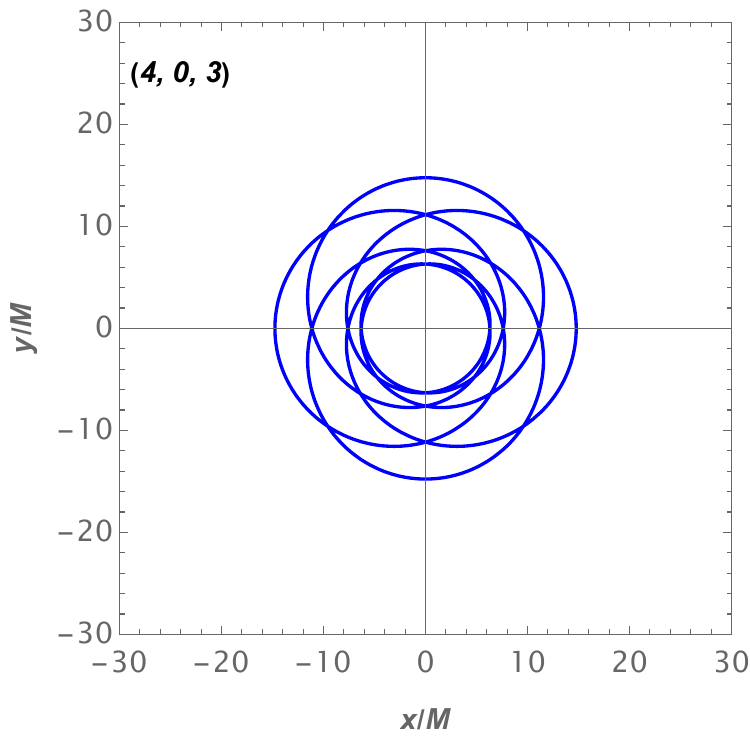} \hfill
    \includegraphics[width=0.32\textwidth]{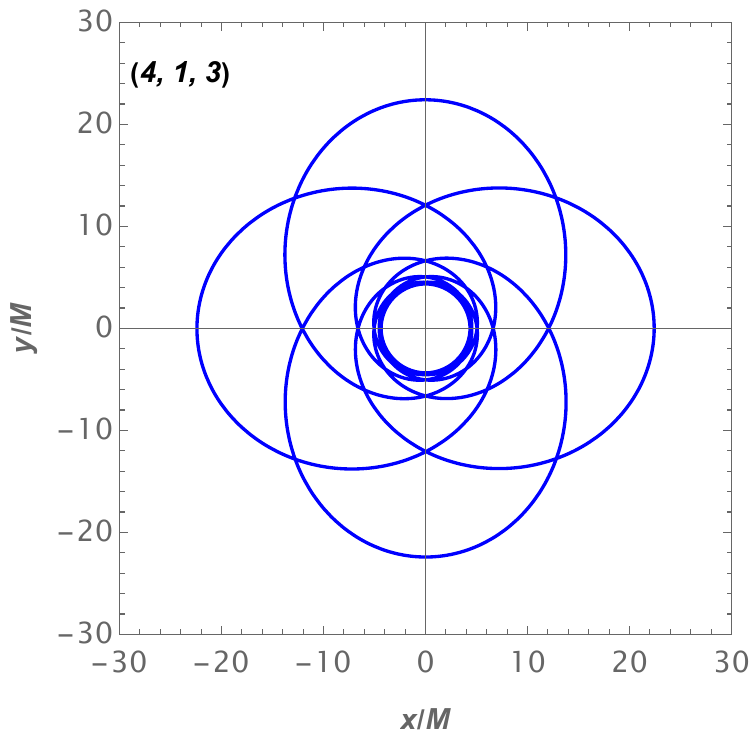} \hfill
    \includegraphics[width=0.32\textwidth]{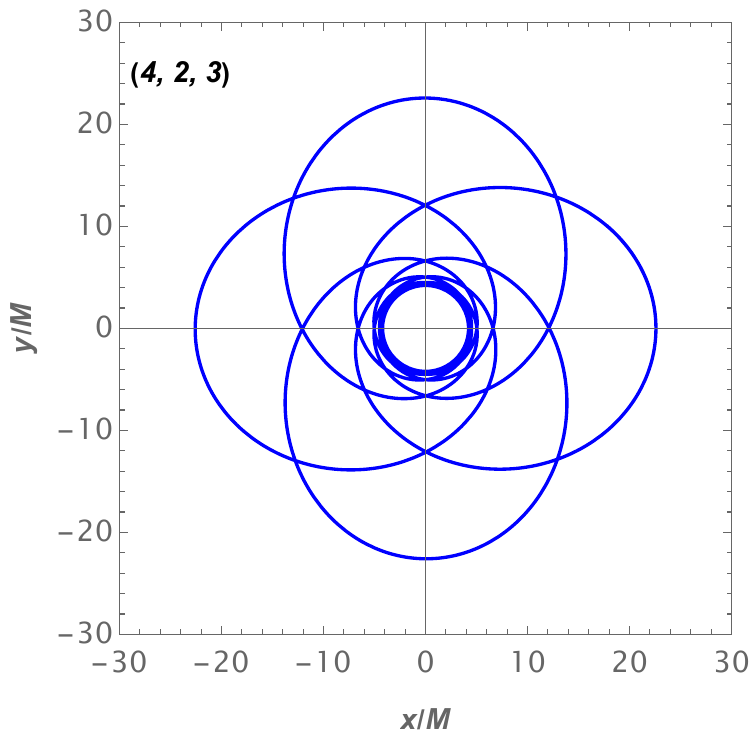}
    
    \caption{The figure illustrates the periodic orbits characterized by different $(z,w,v)$ around the MCBH. Here, $q=0.4$ and  $L_{av}=\frac{1}{2}(L_{MBO}+L_{ISCO})$.}
    \label{fig:periodic}
\end{figure*}

This section is devoted to the periodic orbits around the MCBH. The fundamental frequencies are connected through rational ratios. Note that every periodic orbit is characterized by three integers, which are ($z,w,v$), the zoom, whirl, and vertex numbers, respectively. One can write the rational number in the following form~\cite{Levin_2008} 
\begin{equation}
    g=\frac{\omega_{\phi}}{\omega_r}-1=w+\frac{v}{z}\, ,
\end{equation}
where  $\omega_r$ and $\omega_{\phi}$ refer to the radial and angular frequencies, respectively. It can be rewritten by considering the equations of motion as follows~\cite{2025JCAP...01..091Y, SHABBIR2025101816,Jiang2024PDU}
\begin{equation}
    g=\frac{1}{\pi}\int^{r_2}_{r_1} \frac{\dot{\phi}}{\dot{r}}-1=\frac{1}{\pi} \int^{r_2}_{r_1}\frac{1}{r^2\sqrt{E^2-(1-\frac{2M}{r} e^{\frac{-q^2}{2Mr}})}}\, ,
\end{equation}
where $r_1$ and $r_2$ are the radii of periapsis and apoapsis of the periodic orbits, respectively. In Fig.~\ref{g}, the rational number as a function of the energy and orbital angular momentum is demonstrated. As can be observed from the left panel of this figure, the values of the rational number $g$ increase with the increase of the energy $E$, and there is a sharp rise as $E$ approaches its maximum value. Also, under the influence of the magnetic charge parameter, the values of the rational number $g$ shift toward the lower side of the energy. Similarly, the values of the rational number $g$ shift toward the lower side of the orbital angular momentum. The values of the rational number $g$ decrease with the increase of the orbital angular momentum $L$, and a sharp increase is observed as the angular momentum $L$ approaches its minimum values. Furthermore, the values of the particle energy along various periodic orbits characterized by ($z,w,v$) are numerically calculated in the Table ~\ref{table1} for constant values of the orbital angular momentum $L_{av}=\frac{1}{2}(L_{MBO}+L_{ISCO})$. In Fig.~\ref{fig:periodic}, we plot the periodic orbits around the MCBH by utilizing the results in Table~\ref{table1}. It can be observed from this figure that periodic orbits with high zoom number $z$ reveal more complex and detailed patterns, while those with a larger whirl number $w$ make loops around the MCBH before reaching their farthest point again.

\section{Numerical kludge gravitational waveforms from periodic orbits} \label{Sec:IV}

\begin{figure*}[htbp]
\begin{subfigure}[b]{0.45\textwidth}
\includegraphics[width=\textwidth]{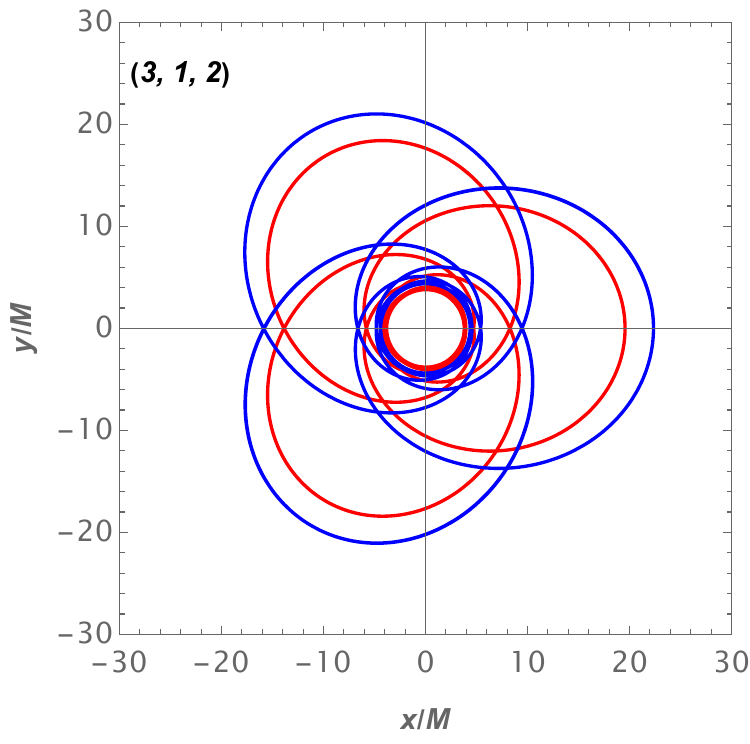}
\end{subfigure}
\begin{subfigure}[b]{0.52\textwidth}
\includegraphics[width=\textwidth]{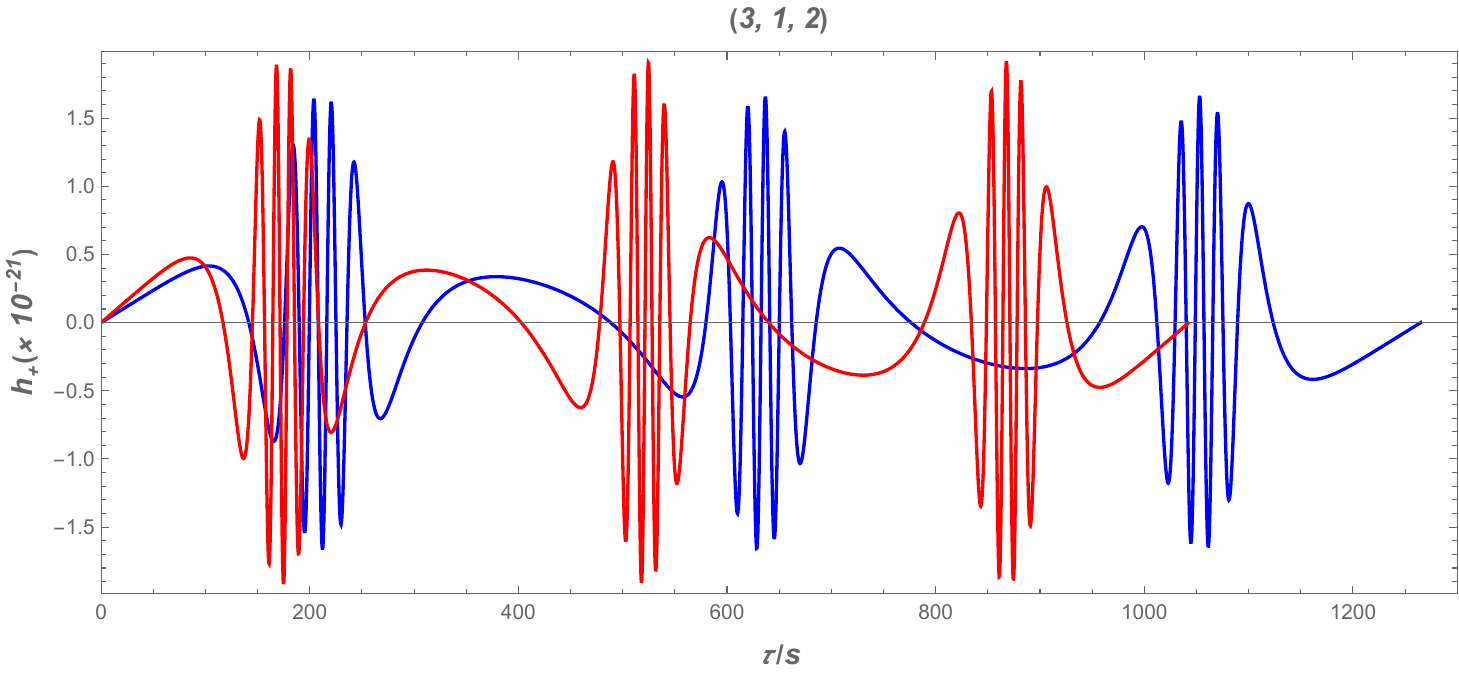}
\includegraphics[width=\textwidth]{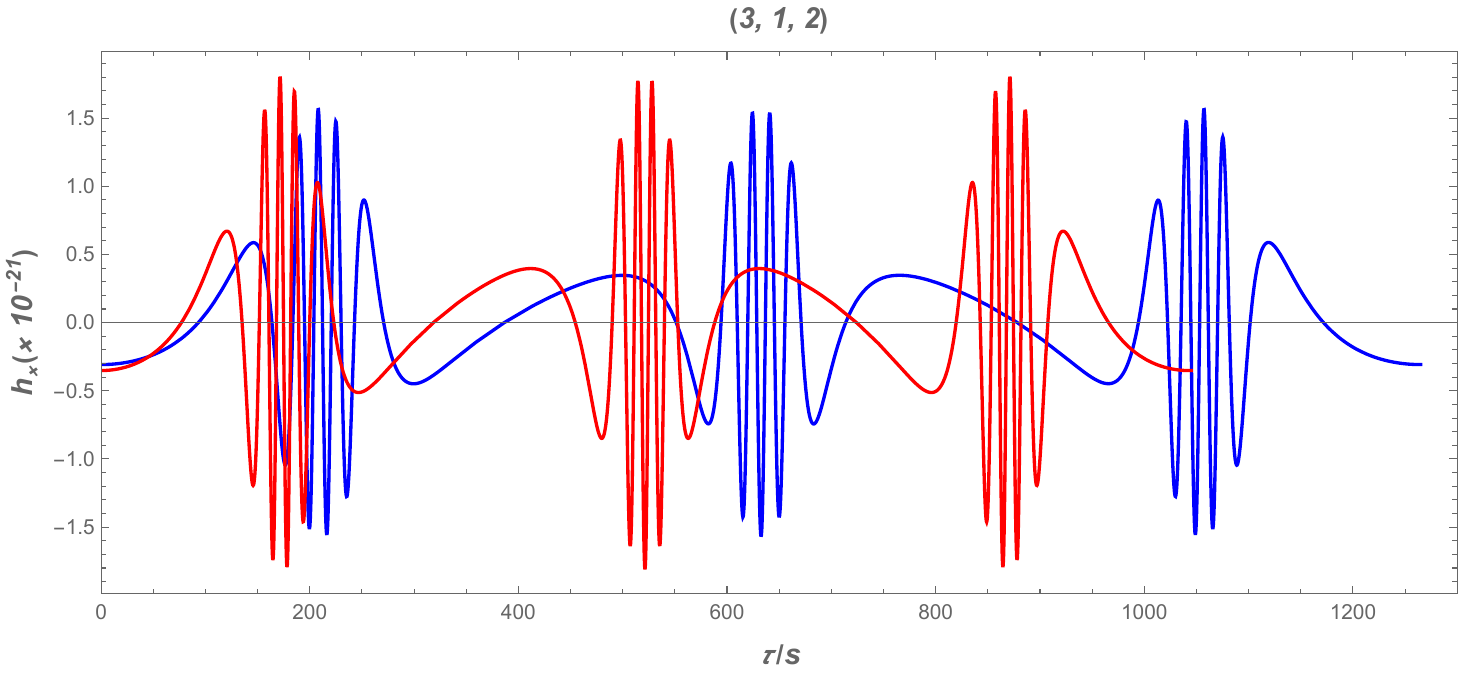}    
\end{subfigure} 
\caption{The plot illustrates the $(3,1,2)$ periodic orbit and its associated GW signal of the EMRI system, which contains of the supermassive MCBH and the small object around this BH. The mass of the SMBH and the small object are set to $M\sim 10^7 M_{\odot}$ and $m\sim 10M_{\odot}$, respectively. The red line represents $q=0.8$, while the blue line corresponds to $q=0.4$.}
\label{fig:combination}
\end{figure*}
\begin{figure*}[htbp]
\begin{subfigure}[b]{0.45\textwidth}
\includegraphics[width=\textwidth]{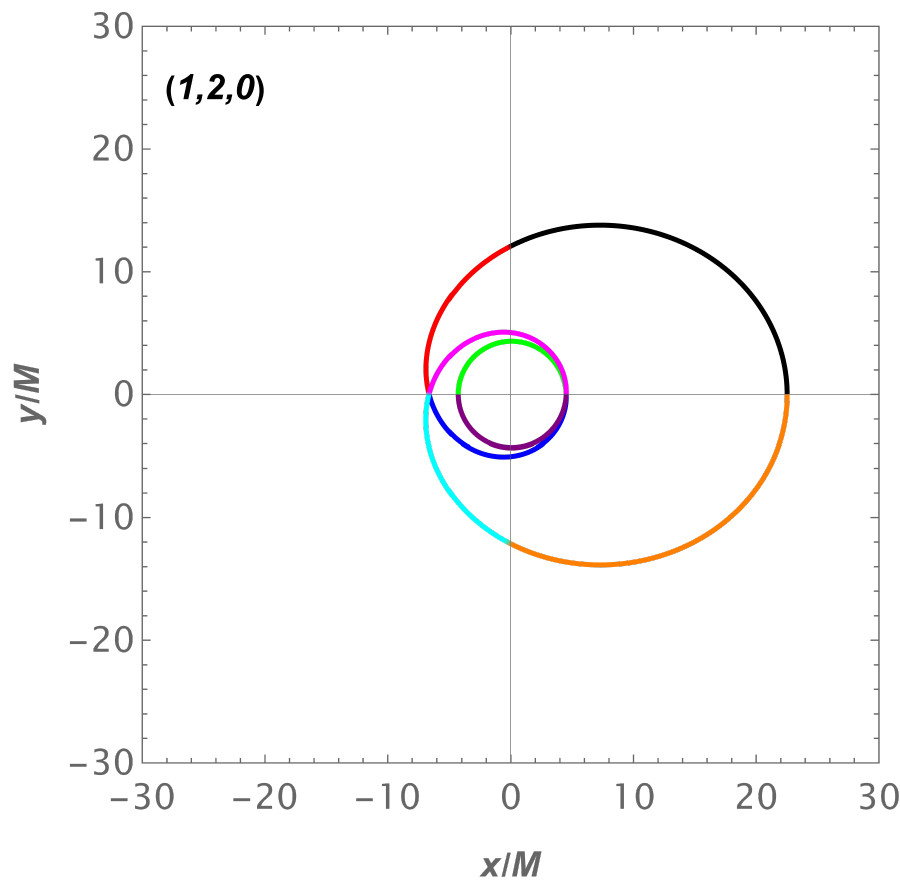}
\end{subfigure}
\begin{subfigure}[b]{0.52\textwidth}
\includegraphics[width=\textwidth]{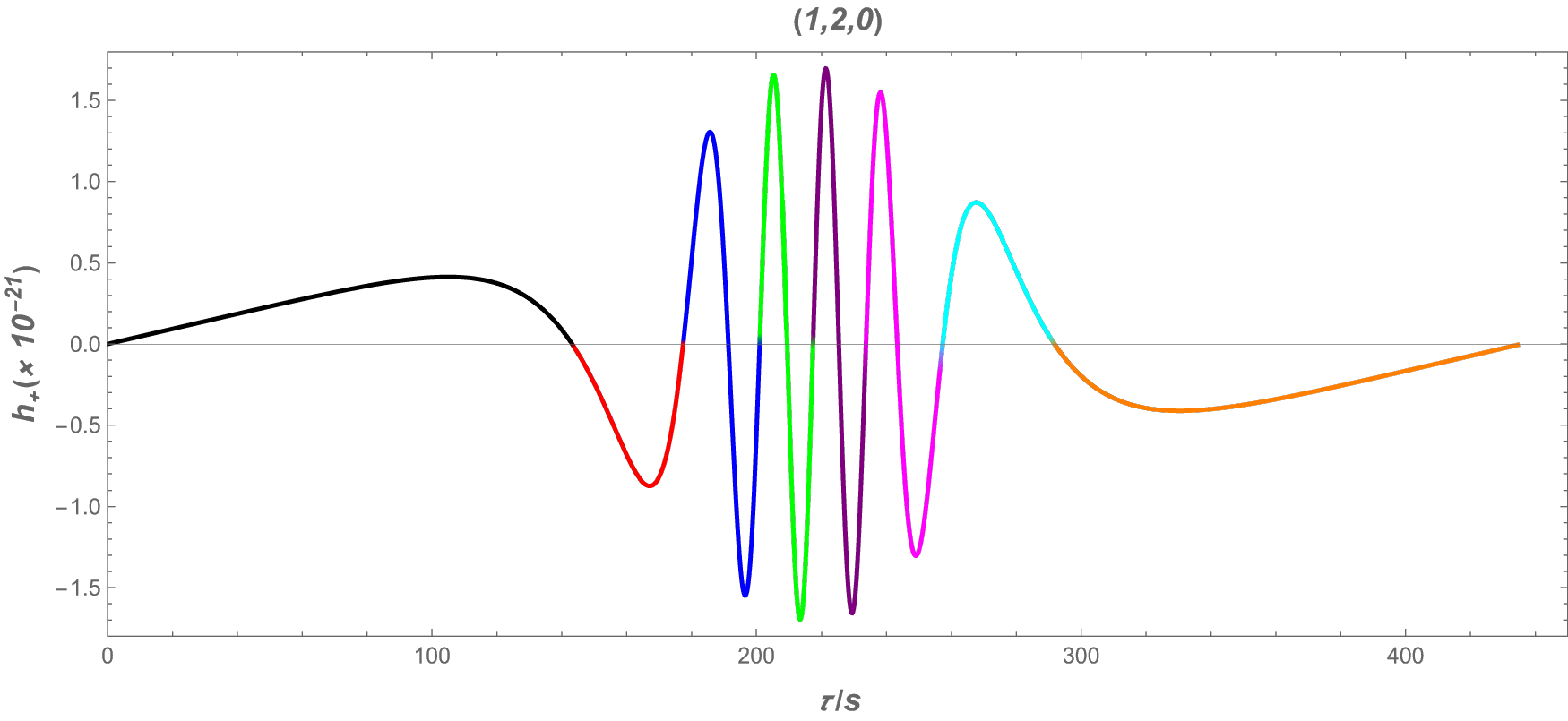}
\includegraphics[width=\textwidth]{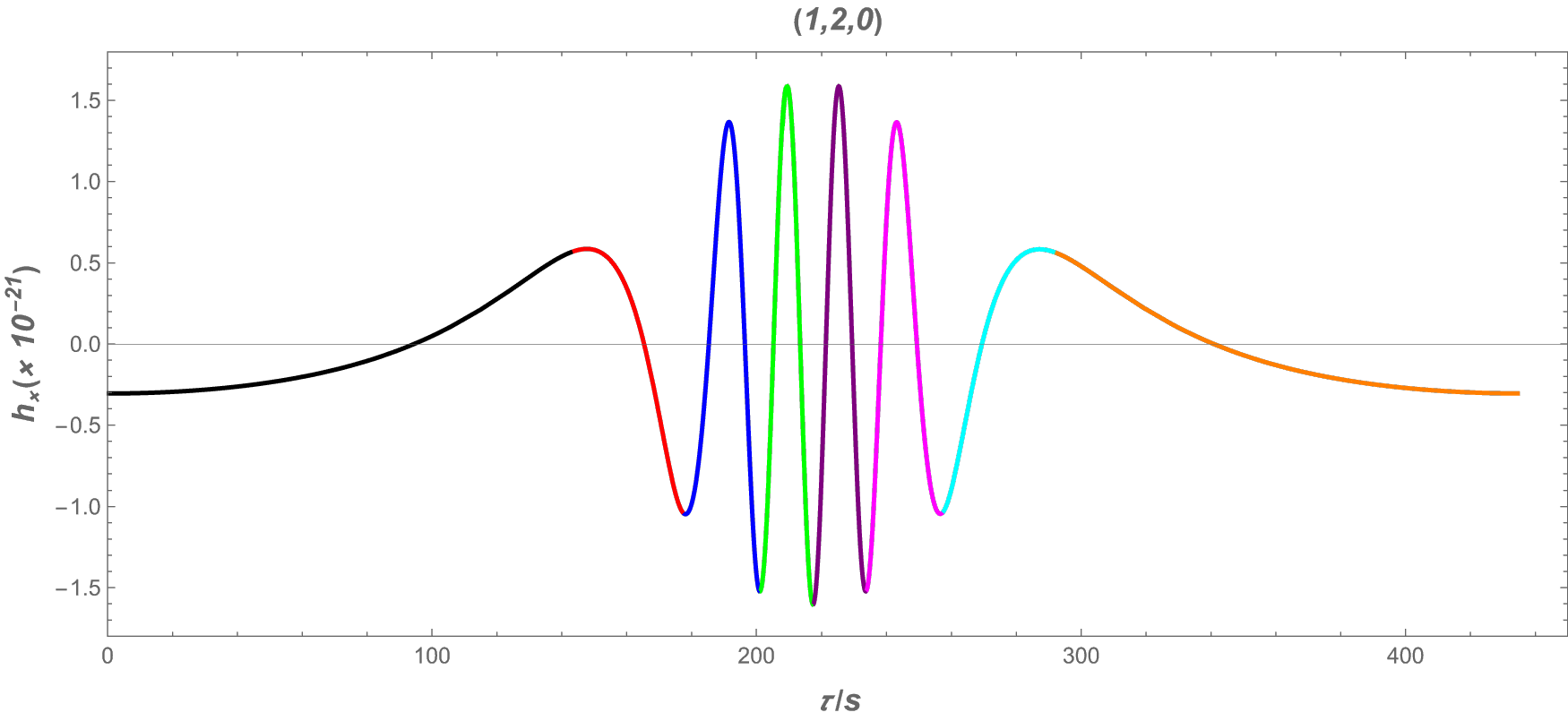} 
\end{subfigure} 
\caption{The plot illustrates the $(1,2,0)$ periodic orbit and its associated GW signal of the EMRI system, which contains of the supermassive MCBH and the small object around this BH. The mass of the SMBH and the small object are set to $M\sim 10^7 M_{\odot}$ and $m\sim 10M_{\odot}$, respectively. Here, each segment is highlighted through different colors.} 
\label{fig:combination2}
\end{figure*}
\begin{figure*}[htbp]
\includegraphics[width=0.85\textwidth]{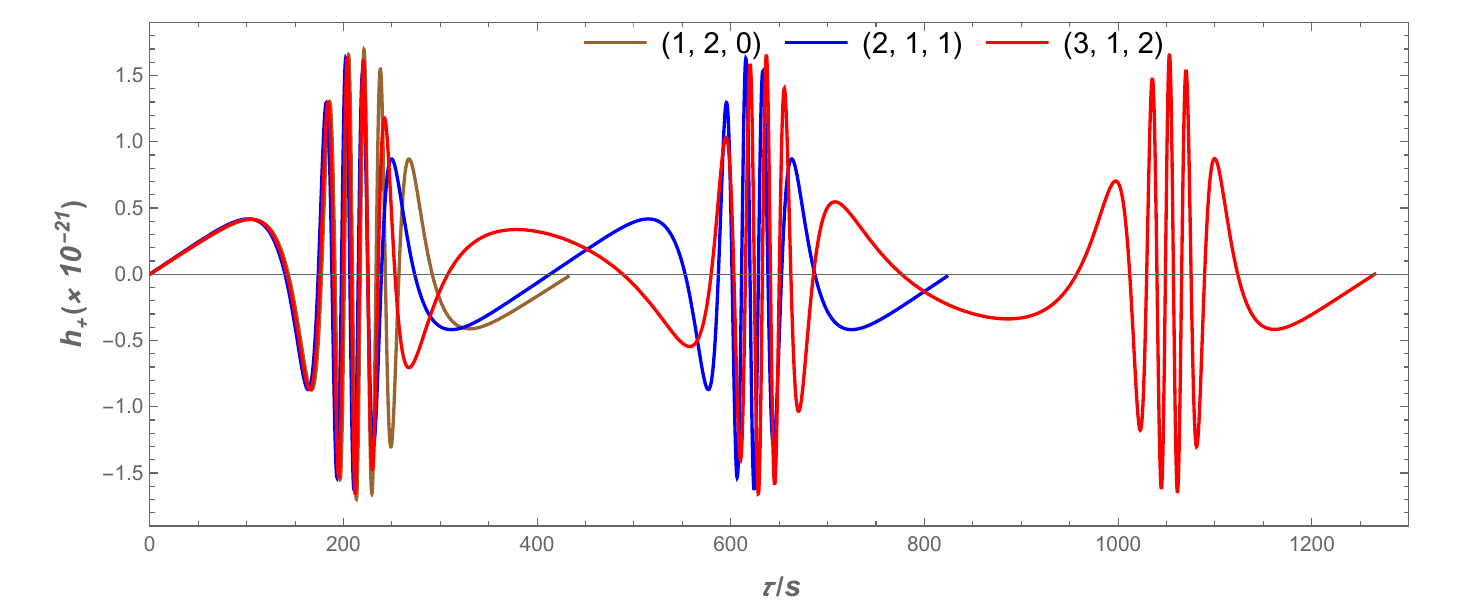}
\includegraphics[width=0.85\textwidth]{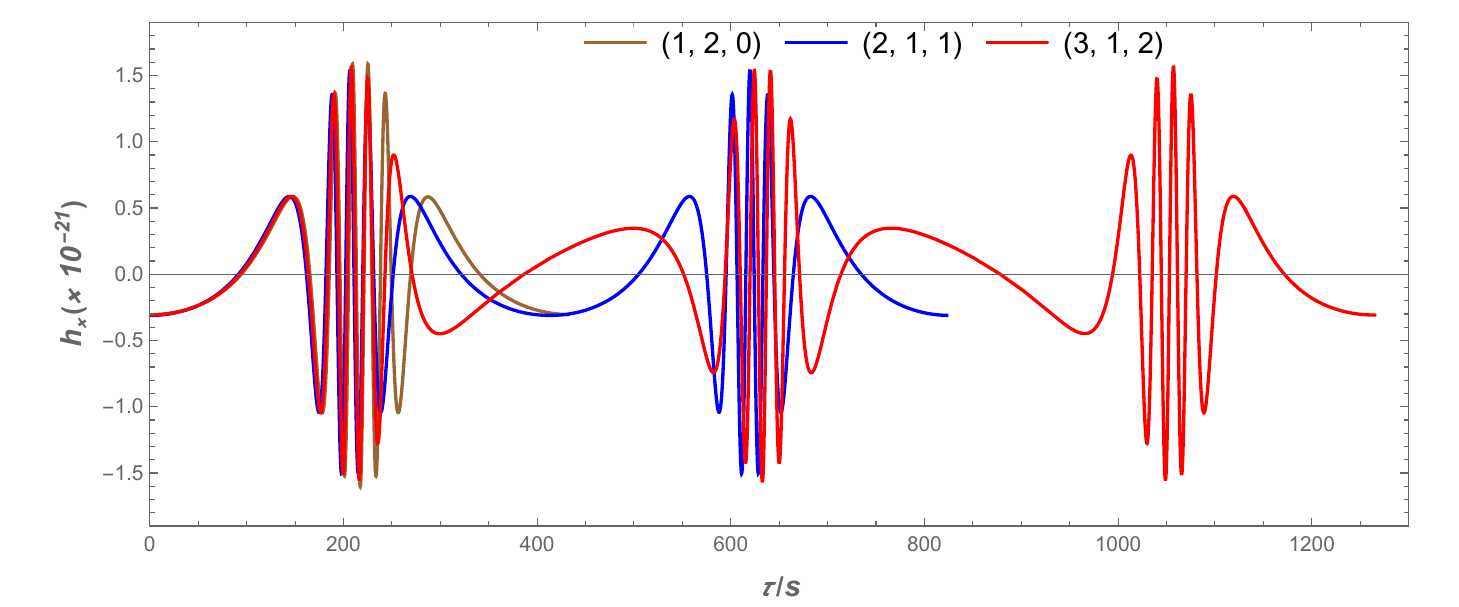}
\caption{The plot shows the contrasting gravitational waveforms produced by different periodic orbits. The upper panel displays the $h_{+}$, while lower panel shows $h_{\times}$. The mass of the SMBH and the small object are set to $M\sim 10^7 M_{\odot}$ and $m\sim 10M_{\odot}$, respectively. Here, we set $q=0.4$.}
\label{fig:GW}
\end{figure*}

In this part, the gravitational waveforms from the periodic orbits around the MCBH are explored by considering the EMRI system. Note that the EMRI system involves a stellar-mass object along the periodic orbit around a supermassive MCBH. This motion emits GWs that may contain information about the supermassive MCBH. In the short term, the energy and orbital angular momentum remain nearly constant, as the small object's orbit evolves over a timescale much longer than its orbital period. As an outcome, the orbit can be approximated as a geodesic on short timescales, allowing us to model the GWs emitted during one full orbital period. It should be noted that the effect of gravitational radiation on the motion of a small celestial body can be neglected over a given short time interval. The gravitational waveforms from the EMRI system can be explored using the numerical kludge waveform model. Firstly, we solve the equations of motion to study the periodic orbits around the MCBH. Subsequently, the gravitational waveforms can be generated through the symmetric and trace-free (STF) mass quadrupole equation of gravitational radiation~\cite{2025JCAP...01..091Y}. One can write its expression for a small object in the following form~\cite{RevModPhys.52.299}
\begin{equation}
I^{ij}=\Big[\int d^3 xx^ix^jT^{tt}(t,x^i)\Big]^{(\text{STF})}    
\end{equation}
with the $tt$-component of the stress-energy tensor for the small celestial body with trajectory $Z^i(t)$ is
\begin{equation}
T^{tt}(t,x^i)=m\delta^3(x^i-Z^i(t)) .
\end{equation}
\begin{figure*}[htbp]
\includegraphics[scale=0.6]{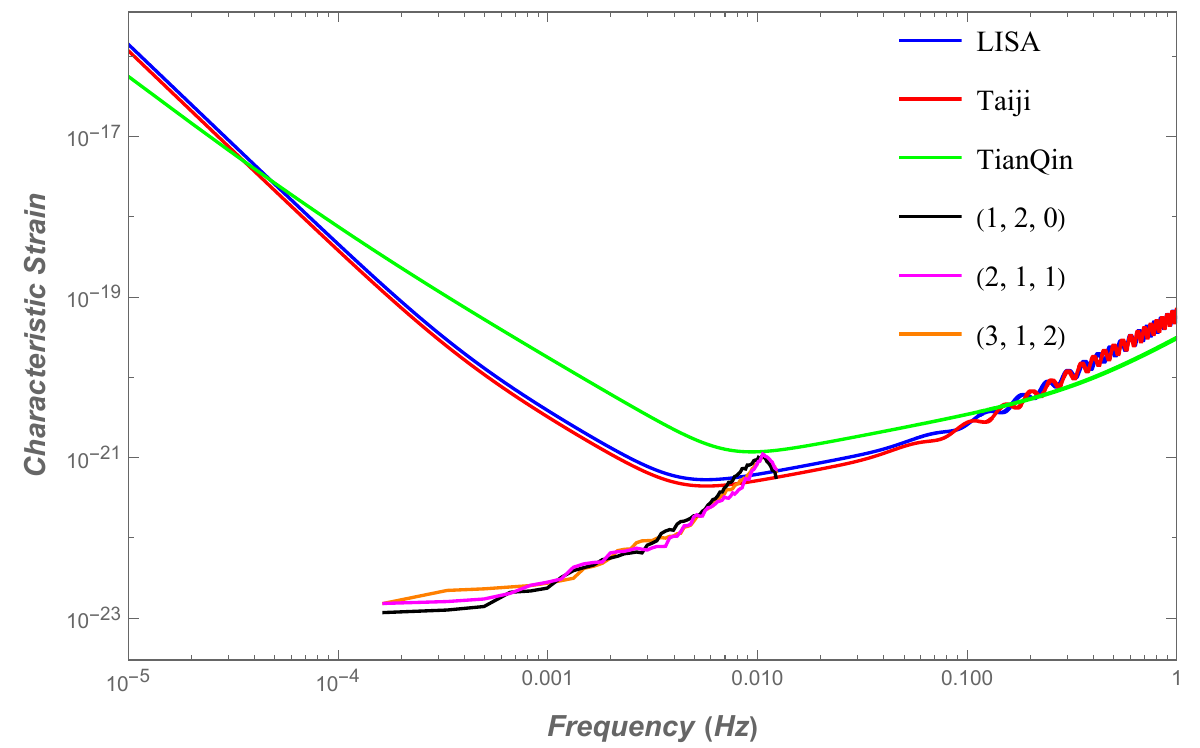}
\caption{The plot demonstrates the comparison of the characteristic strains of gravitational waveforms of periodic orbits with the sensitivity curves of the space-based detectors.}
\label{comparison}
\end{figure*}

By treating the Boyer-Lindquist coordinates as a fictitious spherical polar coordinate system, we project the trajectory of the small object into the Cartesian coordinate system
\begin{align}
x=r\sin{\theta}\cos{\phi}, \quad y=r\sin{\theta}\sin{\phi}, \quad z=r\cos{\theta} .
\end{align}
We can obtain the metric perturbations by considering the above equations as follows
\begin{equation}
h_{ij}=\frac{2}{D_L}\frac{d^2I_{ij}}{dt^2}=\frac{2m}{D_L}(a_ix_j+a_jx_i+2v_iv_j),
\end{equation}
where $a_i$ and $v_i$ refer to the acceleration and velocity of the small object, respectively. $D_L$ represents the luminosity distance from the EMRI system to the detector. This waveform shows the fundamental distribution properties of GW in space. To match real detectors, the calculated GW signal must be modified to fit the detector’s reference system. The GW must be projected into a coordinate system aligned with the detector's framework~\cite{Poisson_Will_2014,2025JCAP...01..091Y,2025EPJC...85...36Z,2024arXiv241101858M}, where the basis takes the form: 
\begin{equation}\label{coordsystem}
\begin{array}{l}
e_{X}=[\cos\zeta,-\sin\zeta,0]\,,\\
e_{Y}=[\cos\iota\sin\zeta,\cos\iota\cos\zeta,-\sin\iota]\,,\\
e_{Z}=[\sin\iota\sin\zeta,\sin\iota\cos\zeta,\cos\iota]\,.
\end{array}
\end{equation}
where $\zeta$ and $\iota$ refer to the longitude of the periastron and inclination angle between the orbit of the test particle and the observation direction, while $e_X$, $e_Y$, $e_Z$ are the orthogonal coordinate basis compatible with the detector. Then we can obtain the corresponding GW polarizations in the following form~\cite{2025JCAP...01..091Y}
\begin{eqnarray}
 h_{+}&=&\frac{1}{2}(e^i_X e^j_X-e^i_Ye^j_Y)h_{ij}, \\
 h_{\times}&=&\frac{1}{2}(e^i_Xe^j_Y+e^i_Ye^j_X)h_{ij}.
\end{eqnarray}
One can rewrite the above equation by considering Eq.~(\ref{coordsystem}) as
\begin{eqnarray}
h_{+}&=&\frac{1}{2}(h_{\zeta\zeta}-h_{\iota\iota}), \\
h_{\times}&=& h_{\iota\zeta}.
\end{eqnarray}
where
\begin{widetext}
\begin{eqnarray}
h_{\zeta\zeta}&=&h_{xx}\cos^2{\zeta}-h_{xy}\sin{2\zeta}+h_{yy}\sin^2{\zeta}, \\
h_{\iota\iota}&=&\cos{\iota}[h_{xx}\sin^2{\zeta}+h_{xy}\sin{2\zeta}+h_{yy}\cos^2{\zeta}]+h_{zz}\sin^2{\iota}-\sin{2\iota}[h_{xz}\sin{\zeta}+h_{yz}\cos{\zeta}], \\
h_{\iota\zeta}&=&\cos{\iota}\Big[\frac{1}{2}h_{xx}\sin{2\zeta}+h_{xy}\cos{2\zeta}-\frac{1}{2}h_{yy}\sin{2\zeta}\Big]+\sin{\iota}[h_{yz}\sin{\zeta}-h_{xz}\cos{\zeta}].
\end{eqnarray}
\end{widetext}

To provide more information, we visualize the GWs that are radiated by the celestial bodies along the periodic orbits around the BHs. In other words, we consider the EMRI system and assume the parameters of this system as: the mass of the celestial body is $m \sim 10 M_{\odot}$, and the mass of the SMBH is $M \sim 10^{7}M_{\odot}$. The latitude and the inclination angle are set as $\zeta=\iota=\pi/4$. We also set the luminosity distance as $D_{L}=200 \text{Mpc}$. The gravitational waveforms ($h_{+}$ and $h_{\times}$) radiated through the $(3,1,2)$ periodic orbit are demonstrated in Fig.~\ref{fig:combination}. We can see from this figure that there are zoom and whirl stages. The calm and rapid oscillations parts correspond to the zoom and whirl stages, respectively. When the small object enters the elliptical orbit far from the SMBH, the GW is radiated as the calm part. In contrast, the GW is radiated as the whirl stage when the small object begins a whirl motion near the SMBH. Note that the frequency of the GW increases dramatically in the whirl stage, and it leads to the intense oscillations. The Fig.~\ref{fig:combination} also shows the difference between $q=0.4$ and $q=0.8$ cases. The blue line corresponds to the $q=0.4$ case, while the red line shows the $q=0.8$ case. One can see from this figure that the zoom and whirl areas shrink with the increase of the magnetic charge parameter $q$. As a result, the gravitational waveforms shift toward smaller values of time, and the values of the polarization components slightly increase with the increase of the magnetic charge parameter. To be more informative, we plot the $(1,2,0)$ periodic orbit and the corresponding gravitational waveforms using the different colors in Fig~\ref{fig:combination2}. From this figure, we can easily see which part of the periodic orbit corresponds to which part of the gravitational waveforms. Moreover, we compare the gravitational waveforms emitted by the different periodic orbits in Fig.~\ref{fig:GW}. We emphasize that our analysis focuses solely on a single orbital period of the small object around the SMBH. Consequently, each periodic orbit generates a distinct gravitational waveform signature.

What is more, one can compare our results with the sensitivity of space-based GW detectors. Note that GWs produced by EMRIs with periodic orbits typically have characteristic frequencies in the mHz range, making them ideal for detection by space-based GW detectors. For doing that, we need to calculate the corresponding characteristic strain by~\cite{2025JCAP...01..091Y}
\begin{equation}
h_c(f)=2f\Big(|\Tilde{h}_{+}(f)|^2+|\Tilde{h}_{\times}(f)|^2\Big)^{1/2}
\end{equation}
where $\Tilde{h}_{+,\times}(f)$ refers to the frequency spectra, which is obtained by performing discrete Fourier transforms on the time-domain gravitational waveforms. 
We plot the characteristic strain for different periodic orbits together with the sensitivity curves of some space-based detectors in Fig.~\ref{comparison}. It can be seen from this figure that some values of the characteristic strain are above the sensitivity curve of the space-based detectors. In other words, the GW emitted from EMRIs with periodic orbits is potentially detectable through these space-based detectors.

\section{Conclusion}\label{Sec:V}

In this paper, we investigated the dynamics of test particles around a magnetically charged black hole (MCBH). Using Lagrangian formalism, we derived the effective potential and analyzed its behavior. Our results demonstrated a slight shift to the left in the potential curves, indicating a decrease in the radial distance $r$ as the magnetic charge parameter $q$ and the orbital angular momentum increase. The potential maximum rises, reflecting a stronger gravitational barrier due to the influence of the magnetic charge parameter. We further determined the radii of marginally bound orbits (MBOs) and innermost stable circular orbits (ISCOs) under the required conditions. Our analysis shows that the magnetic charge parameter $q$ reduces not only these orbital radii but also the corresponding values of the specific angular momentum and energy.

Furthermore, we analyzed the influence of the magnetic charge parameter $q$ on the rational number $g$ that characterizes the relationship between energy $E$ and angular momentum $L$, as illustrated in Fig.~\ref{g}. As the magnetic charge parameter $q$ increases, the rational number curves shift to the left toward lower values of $E$ and $L$, indicating a reduction in both quantities due to the magnetic charge parameter. We further explored this behavior numerically by examining the relationship between $g$ and the averaged angular momentum $L_{av}=\frac{1}{2}(L_{MBO}+L_{ISCO})$ for the corresponding periodic orbits characterized by three integers such as zoom $z$, whirl $w$, and vertex $v$ numbers, providing a comprehensive mapping of the dynamics under varying the magnetic charge parameter $q$; see Table~\ref{table1}. 

Finally, we analyzed the gravitational waveforms produced by periodic orbits in the MCBH spacetime. We modeled the inspiral of a stellar-mass object ($m \sim 10 M_{\odot}$), as a timelike particle orbiting a SMBH with mass $M \sim 10^7 M_{\odot}$, where the trajectory is governed by periodic orbits. Using the numerical kludge method, we computed the gravitational waveforms for extreme mass-ratio inspirals (EMRIs), focusing on the $(3,1,2)$ orbit as a representative case, as shown in Fig.~\ref{fig:combination}. Our results demonstrated that the magnetic charge parameter $q$ significantly modifies the zoom-whirl dynamics, leading to notable changes in the waveforms ($h_{+}$ and $h_{\times}$). To further elucidate these effects, we analyze the gravitational waveforms emitted from the (1,2,0) orbit (see Fig.~\ref{fig:combination2}), highlighting how specific orbital features produce the waveform structure. Additionally, we presented a broader comparison of waveforms from different periodic orbits, showing the resulting gravitational waveforms, as illustrated in Fig.~\ref{fig:GW}.

We also provided a more appropriate analysis associated with a better fit of the observational data from future space-based detectors such as LISA, Taiji, and TianQin \cite{Amaro-Seoane2017LISA, Hu:10.1093/nsr/nwx116, Luo:TianQin2016}. We showed that our findings related to gravitational waveforms emitted from EMRIs through periodic orbits satisfy the observational data and are potentially detectable through these space-based detectors. The results are astrophysically significant, especially considering the importance of gravitational waveforms from EMRIs. These findings emphasize the role of the magnetic charge in altering EMRI waveform signatures, which may have observable implications for future GW detectors.

\section*{Acknowledgments}

SS is supported by the National Natural Science Foundation of China under Grant No. W2433018. TZ is supported by the National Key Research and Development Program of China under Grant No. 2020YFC2201503, the National Natural Science Foundation of China under Grants No.~12275238 and No.~11675143, the Zhejiang Provincial Natural Science Foundation of China under Grants No.~LR21A050001 and No.~LY20A050002, and the Fundamental Research Funds for the Provincial Universities of Zhejiang in China under Grant No.~RF-A2019015.

\bibliographystyle{apsrev4-1}
\bibliography{ref}

\begin{thebibliography}{78}%
\makeatletter
\providecommand \@ifxundefined [1]{%
 \@ifx{#1\undefined}
}%
\providecommand \@ifnum [1]{%
 \ifnum #1\expandafter \@firstoftwo
 \else \expandafter \@secondoftwo
 \fi
}%
\providecommand \@ifx [1]{%
 \ifx #1\expandafter \@firstoftwo
 \else \expandafter \@secondoftwo
 \fi
}%
\providecommand \natexlab [1]{#1}%
\providecommand \enquote  [1]{``#1''}%
\providecommand \bibnamefont  [1]{#1}%
\providecommand \bibfnamefont [1]{#1}%
\providecommand \citenamefont [1]{#1}%
\providecommand \href@noop [0]{\@secondoftwo}%
\providecommand \href [0]{\begingroup \@sanitize@url \@href}%
\providecommand \@href[1]{\@@startlink{#1}\@@href}%
\providecommand \@@href[1]{\endgroup#1\@@endlink}%
\providecommand \@sanitize@url [0]{\catcode `\\12\catcode `\$12\catcode `\&12\catcode `\#12\catcode `\^12\catcode `\_12\catcode `\%12\relax}%
\providecommand \@@startlink[1]{}%
\providecommand \@@endlink[0]{}%
\providecommand \url  [0]{\begingroup\@sanitize@url \@url }%
\providecommand \@url [1]{\endgroup\@href {#1}{\urlprefix }}%
\providecommand \urlprefix  [0]{URL }%
\providecommand \Eprint [0]{\href }%
\providecommand \doibase [0]{http://dx.doi.org/}%
\providecommand \selectlanguage [0]{\@gobble}%
\providecommand \bibinfo  [0]{\@secondoftwo}%
\providecommand \bibfield  [0]{\@secondoftwo}%
\providecommand \translation [1]{[#1]}%
\providecommand \BibitemOpen [0]{}%
\providecommand \bibitemStop [0]{}%
\providecommand \bibitemNoStop [0]{.\EOS\space}%
\providecommand \EOS [0]{\spacefactor3000\relax}%
\providecommand \BibitemShut  [1]{\csname bibitem#1\endcsname}%
\let\auto@bib@innerbib\@empty
\bibitem [{\citenamefont {{Abbott}}\ and\ \citenamefont {et~al. {(Virgo and LIGO Scientific Collaborations)}}(2016{\natexlab{a}})}]{Abbott16a}%
  \BibitemOpen
  \bibfield  {author} {\bibinfo {author} {\bibfnamefont {B.~P.}\ \bibnamefont {{Abbott}}}\ and\ \bibinfo {author} {\bibnamefont {et~al. {(Virgo and LIGO Scientific Collaborations)}}},\ }\href {\doibase 10.1103/PhysRevLett.116.061102} {\bibfield  {journal} {\bibinfo  {journal} {Phys. Rev. Lett.}\ }\textbf {\bibinfo {volume} {116}},\ \bibinfo {eid} {061102} (\bibinfo {year} {2016}{\natexlab{a}})},\ \Eprint {http://arxiv.org/abs/1602.03837} {arXiv:1602.03837 [gr-qc]} \BibitemShut {NoStop}%
\bibitem [{\citenamefont {{Abbott}}\ and\ \citenamefont {et~al. {(Virgo and LIGO Scientific Collaborations)}}(2016{\natexlab{b}})}]{Abbott16b}%
  \BibitemOpen
  \bibfield  {author} {\bibinfo {author} {\bibfnamefont {B.~P.}\ \bibnamefont {{Abbott}}}\ and\ \bibinfo {author} {\bibnamefont {et~al. {(Virgo and LIGO Scientific Collaborations)}}},\ }\href {\doibase 10.1103/PhysRevLett.116.241102} {\bibfield  {journal} {\bibinfo  {journal} {Phys. Rev. Lett.}\ }\textbf {\bibinfo {volume} {116}},\ \bibinfo {eid} {241102} (\bibinfo {year} {2016}{\natexlab{b}})},\ \Eprint {http://arxiv.org/abs/1602.03840} {arXiv:1602.03840 [gr-qc]} \BibitemShut {NoStop}%
\bibitem [{\citenamefont {{Akiyama}}\ and\ \citenamefont {et~al. {(Event Horizon Telescope Collaboration)}}(2019)}]{Akiyama19L1}%
  \BibitemOpen
  \bibfield  {author} {\bibinfo {author} {\bibfnamefont {K.}~\bibnamefont {{Akiyama}}}\ and\ \bibinfo {author} {\bibnamefont {et~al. {(Event Horizon Telescope Collaboration)}}},\ }\href {\doibase 10.3847/2041-8213/ab0ec7} {\bibfield  {journal} {\bibinfo  {journal} {Astrophys. J.}\ }\textbf {\bibinfo {volume} {875}},\ \bibinfo {eid} {L1} (\bibinfo {year} {2019})},\ \Eprint {http://arxiv.org/abs/1906.11238} {arXiv:1906.11238 [astro-ph.GA]} \BibitemShut {NoStop}%
\bibitem [{\citenamefont {{Akiyama}}\ and\ \citenamefont {et~al. {(Event Horizon Telescope Collaboration)}}(2022)}]{Akiyama22L12}%
  \BibitemOpen
  \bibfield  {author} {\bibinfo {author} {\bibfnamefont {K.}~\bibnamefont {{Akiyama}}}\ and\ \bibinfo {author} {\bibnamefont {et~al. {(Event Horizon Telescope Collaboration)}}},\ }\href {\doibase 10.3847/2041-8213/ac6674} {\bibfield  {journal} {\bibinfo  {journal} {Astrophys. J. Lett.}\ }\textbf {\bibinfo {volume} {930}},\ \bibinfo {eid} {L12} (\bibinfo {year} {2022})}\BibitemShut {NoStop}%
\bibitem [{\citenamefont {Will}(2014)}]{Will14LRR}%
  \BibitemOpen
  \bibfield  {author} {\bibinfo {author} {\bibfnamefont {C.~M.}\ \bibnamefont {Will}},\ }\href {\doibase 10.12942/lrr-2014-4} {\bibfield  {journal} {\bibinfo  {journal} {Living Rev. Rel.}\ }\textbf {\bibinfo {volume} {17}},\ \bibinfo {pages} {4} (\bibinfo {year} {2014})},\ \Eprint {http://arxiv.org/abs/1403.7377} {arXiv:1403.7377 [gr-qc]} \BibitemShut {NoStop}%
\bibitem [{\citenamefont {Psaltis}\ \emph {et~al.}(2020)\citenamefont {Psaltis} \emph {et~al.}}]{Psaltis20PRL}%
  \BibitemOpen
  \bibfield  {author} {\bibinfo {author} {\bibfnamefont {D.}~\bibnamefont {Psaltis}} \emph {et~al.} (\bibinfo {collaboration} {Event Horizon Telescope}),\ }\href {\doibase 10.1103/PhysRevLett.125.141104} {\bibfield  {journal} {\bibinfo  {journal} {Phys. Rev. Lett.}\ }\textbf {\bibinfo {volume} {125}},\ \bibinfo {pages} {141104} (\bibinfo {year} {2020})},\ \Eprint {http://arxiv.org/abs/2010.01055} {arXiv:2010.01055 [gr-qc]} \BibitemShut {NoStop}%
\bibitem [{\citenamefont {Hawking}\ and\ \citenamefont {Penrose}(1970)}]{Hawking1970}%
  \BibitemOpen
  \bibfield  {author} {\bibinfo {author} {\bibfnamefont {S.~W.}\ \bibnamefont {Hawking}}\ and\ \bibinfo {author} {\bibfnamefont {R.}~\bibnamefont {Penrose}},\ }\href {\doibase 10.1098/rspa.1970.0021} {\bibfield  {journal} {\bibinfo  {journal} {Proc. Roy. Soc. Lond. A}\ }\textbf {\bibinfo {volume} {314}},\ \bibinfo {pages} {529} (\bibinfo {year} {1970})}\BibitemShut {NoStop}%
\bibitem [{\citenamefont {Bardeen}(1968)}]{Bardeen68}%
  \BibitemOpen
  \bibfield  {author} {\bibinfo {author} {\bibfnamefont {J.}~\bibnamefont {Bardeen}},\ }in\ \href@noop {} {\emph {\bibinfo {booktitle} {Proceedings of GR5}}},\ \bibinfo {editor} {edited by\ \bibinfo {editor} {\bibfnamefont {C.}~\bibnamefont {DeWitt}}\ and\ \bibinfo {editor} {\bibfnamefont {B.}~\bibnamefont {DeWitt}}},\ \bibinfo {organization} {Tbilisi, USSR}\ (\bibinfo  {publisher} {Gordon and Breach},\ \bibinfo {year} {1968})\ p.\ \bibinfo {pages} {174}\BibitemShut {NoStop}%
\bibitem [{\citenamefont {Ay\'on-Beato}\ and\ \citenamefont {Garc\'{\i}a}(1998)}]{Eyon}%
  \BibitemOpen
  \bibfield  {author} {\bibinfo {author} {\bibfnamefont {E.}~\bibnamefont {Ay\'on-Beato}}\ and\ \bibinfo {author} {\bibfnamefont {A.}~\bibnamefont {Garc\'{\i}a}},\ }\href {\doibase 10.1103/PhysRevLett.80.5056} {\bibfield  {journal} {\bibinfo  {journal} {Phys. Rev. Lett.}\ }\textbf {\bibinfo {volume} {80}},\ \bibinfo {pages} {5056} (\bibinfo {year} {1998})}\BibitemShut {NoStop}%
\bibitem [{\citenamefont {{Hayward}}(2006)}]{Hayward06}%
  \BibitemOpen
  \bibfield  {author} {\bibinfo {author} {\bibfnamefont {S.~A.}\ \bibnamefont {{Hayward}}},\ }\href {\doibase 10.1103/PhysRevLett.96.031103} {\bibfield  {journal} {\bibinfo  {journal} {Phys. Rev. Lett.}\ }\textbf {\bibinfo {volume} {96}},\ \bibinfo {eid} {031103} (\bibinfo {year} {2006})},\ \Eprint {http://arxiv.org/abs/gr-qc/0506126} {gr-qc/0506126} \BibitemShut {NoStop}%
\bibitem [{\citenamefont {Ayon-Beato}\ and\ \citenamefont {Garcia}(2000)}]{Ayon-Beato2000PLB}%
  \BibitemOpen
  \bibfield  {author} {\bibinfo {author} {\bibfnamefont {E.}~\bibnamefont {Ayon-Beato}}\ and\ \bibinfo {author} {\bibfnamefont {A.}~\bibnamefont {Garcia}},\ }\href {\doibase 10.1016/S0370-2693(00)01125-4} {\bibfield  {journal} {\bibinfo  {journal} {Phys. Lett. B}\ }\textbf {\bibinfo {volume} {493}},\ \bibinfo {pages} {149} (\bibinfo {year} {2000})},\ \Eprint {http://arxiv.org/abs/gr-qc/0009077} {arXiv:gr-qc/0009077} \BibitemShut {NoStop}%
\bibitem [{\citenamefont {Bronnikov}(2001)}]{Bronnikov2000}%
  \BibitemOpen
  \bibfield  {author} {\bibinfo {author} {\bibfnamefont {K.~A.}\ \bibnamefont {Bronnikov}},\ }\href {\doibase 10.1103/PhysRevD.63.044005} {\bibfield  {journal} {\bibinfo  {journal} {Phys. Rev. D}\ }\textbf {\bibinfo {volume} {63}},\ \bibinfo {pages} {044005} (\bibinfo {year} {2001})},\ \Eprint {http://arxiv.org/abs/gr-qc/0006014} {arXiv:gr-qc/0006014} \BibitemShut {NoStop}%
\bibitem [{\citenamefont {{Bardeen}}(1968)}]{Bardeen68B}%
  \BibitemOpen
  \bibfield  {author} {\bibinfo {author} {\bibfnamefont {J.}~\bibnamefont {{Bardeen}}},\ }in\ \href@noop {} {\emph {\bibinfo {booktitle} {Proceedings of the 5th International Conference on Gravitation and the Theory of Relativity}}}\ (\bibinfo {year} {1968})\ p.~\bibinfo {pages} {87}\BibitemShut {NoStop}%
\bibitem [{\citenamefont {Dymnikova}(1992)}]{Dymnikova92}%
  \BibitemOpen
  \bibfield  {author} {\bibinfo {author} {\bibfnamefont {I.}~\bibnamefont {Dymnikova}},\ }\href {\doibase 10.1007/BF00760226} {\bibfield  {journal} {\bibinfo  {journal} {Gen. Rel. Grav.}\ }\textbf {\bibinfo {volume} {24}},\ \bibinfo {pages} {235} (\bibinfo {year} {1992})}\BibitemShut {NoStop}%
\bibitem [{\citenamefont {Balart}\ and\ \citenamefont {Vagenas}(2014)}]{Balart14PLB}%
  \BibitemOpen
  \bibfield  {author} {\bibinfo {author} {\bibfnamefont {L.}~\bibnamefont {Balart}}\ and\ \bibinfo {author} {\bibfnamefont {E.~C.}\ \bibnamefont {Vagenas}},\ }\href {\doibase 10.1016/j.physletb.2014.01.024} {\bibfield  {journal} {\bibinfo  {journal} {Phys. Lett. B}\ }\textbf {\bibinfo {volume} {730}},\ \bibinfo {pages} {14} (\bibinfo {year} {2014})},\ \Eprint {http://arxiv.org/abs/1401.2136} {arXiv:1401.2136 [gr-qc]} \BibitemShut {NoStop}%
\bibitem [{\citenamefont {Kumar}\ \emph {et~al.}(2020)\citenamefont {Kumar}, \citenamefont {Kumar},\ and\ \citenamefont {Ghosh}}]{Kumar20MNRAS}%
  \BibitemOpen
  \bibfield  {author} {\bibinfo {author} {\bibfnamefont {R.}~\bibnamefont {Kumar}}, \bibinfo {author} {\bibfnamefont {A.}~\bibnamefont {Kumar}}, \ and\ \bibinfo {author} {\bibfnamefont {S.~G.}\ \bibnamefont {Ghosh}},\ }\href {\doibase 10.1093/mnras/staa2104} {\bibfield  {journal} {\bibinfo  {journal} {Mon. Not. R. Astron. Soc.}\ }\textbf {\bibinfo {volume} {497}},\ \bibinfo {pages} {3845} (\bibinfo {year} {2020})},\ \Eprint {http://arxiv.org/abs/2006.09869} {arXiv:2006.09869 [gr-qc]} \BibitemShut {NoStop}%
\bibitem [{\citenamefont {Fan}\ and\ \citenamefont {Wang}(2016)}]{Fan16}%
  \BibitemOpen
  \bibfield  {author} {\bibinfo {author} {\bibfnamefont {Z.-Y.}\ \bibnamefont {Fan}}\ and\ \bibinfo {author} {\bibfnamefont {X.}~\bibnamefont {Wang}},\ }\href {\doibase 10.1103/PhysRevD.94.124027} {\bibfield  {journal} {\bibinfo  {journal} {Phys. Rev. D}\ }\textbf {\bibinfo {volume} {94}},\ \bibinfo {pages} {124027} (\bibinfo {year} {2016})},\ \Eprint {http://arxiv.org/abs/1610.02636} {arXiv:1610.02636 [gr-qc]} \BibitemShut {NoStop}%
\bibitem [{\citenamefont {Toshmatov}\ \emph {et~al.}(2014)\citenamefont {Toshmatov}, \citenamefont {Ahmedov}, \citenamefont {Abdujabbarov},\ and\ \citenamefont {Stuchl\'{i}k}}]{Toshmatov17}%
  \BibitemOpen
  \bibfield  {author} {\bibinfo {author} {\bibfnamefont {B.}~\bibnamefont {Toshmatov}}, \bibinfo {author} {\bibfnamefont {B.}~\bibnamefont {Ahmedov}}, \bibinfo {author} {\bibfnamefont {A.}~\bibnamefont {Abdujabbarov}}, \ and\ \bibinfo {author} {\bibfnamefont {Z.}~\bibnamefont {Stuchl\'{i}k}},\ }\href {\doibase 10.1103/PhysRevD.89.104017} {\bibfield  {journal} {\bibinfo  {journal} {Phys. Rev. D}\ }\textbf {\bibinfo {volume} {89}},\ \bibinfo {pages} {104017} (\bibinfo {year} {2014})},\ \Eprint {http://arxiv.org/abs/1404.6443} {arXiv:1404.6443 [gr-qc]} \BibitemShut {NoStop}%
\bibitem [{\citenamefont {Neves}\ and\ \citenamefont {Saa}(2014)}]{Neves14PLB}%
  \BibitemOpen
  \bibfield  {author} {\bibinfo {author} {\bibfnamefont {J.~C.~S.}\ \bibnamefont {Neves}}\ and\ \bibinfo {author} {\bibfnamefont {A.}~\bibnamefont {Saa}},\ }\href {\doibase 10.1016/j.physletb.2014.05.026} {\bibfield  {journal} {\bibinfo  {journal} {Phys. Lett. B}\ }\textbf {\bibinfo {volume} {734}},\ \bibinfo {pages} {44} (\bibinfo {year} {2014})},\ \Eprint {http://arxiv.org/abs/1402.2694} {arXiv:1402.2694 [gr-qc]} \BibitemShut {NoStop}%
\bibitem [{\citenamefont {{Narzilloev}}\ \emph {et~al.}(2020)\citenamefont {{Narzilloev}}, \citenamefont {{Rayimbaev}}, \citenamefont {{Shaymatov}}, \citenamefont {{Abdujabbarov}}, \citenamefont {{Ahmedov}},\ and\ \citenamefont {{Bambi}}}]{Narzilloev20b}%
  \BibitemOpen
  \bibfield  {author} {\bibinfo {author} {\bibfnamefont {B.}~\bibnamefont {{Narzilloev}}}, \bibinfo {author} {\bibfnamefont {J.}~\bibnamefont {{Rayimbaev}}}, \bibinfo {author} {\bibfnamefont {S.}~\bibnamefont {{Shaymatov}}}, \bibinfo {author} {\bibfnamefont {A.}~\bibnamefont {{Abdujabbarov}}}, \bibinfo {author} {\bibfnamefont {B.}~\bibnamefont {{Ahmedov}}}, \ and\ \bibinfo {author} {\bibfnamefont {C.}~\bibnamefont {{Bambi}}},\ }\href {\doibase 10.1103/PhysRevD.102.104062} {\bibfield  {journal} {\bibinfo  {journal} {Phys. Rev. D}\ }\textbf {\bibinfo {volume} {102}},\ \bibinfo {eid} {104062} (\bibinfo {year} {2020})},\ \Eprint {http://arxiv.org/abs/2011.06148} {arXiv:2011.06148 [gr-qc]} \BibitemShut {NoStop}%
\bibitem [{\citenamefont {Ghosh}\ and\ \citenamefont {Walia}(2021)}]{Ghosh21AP}%
  \BibitemOpen
  \bibfield  {author} {\bibinfo {author} {\bibfnamefont {S.~G.}\ \bibnamefont {Ghosh}}\ and\ \bibinfo {author} {\bibfnamefont {R.~K.}\ \bibnamefont {Walia}},\ }\href {\doibase 10.1016/j.aop.2021.168619} {\bibfield  {journal} {\bibinfo  {journal} {Ann. Phys.}\ }\textbf {\bibinfo {volume} {434}},\ \bibinfo {pages} {168619} (\bibinfo {year} {2021})}\BibitemShut {NoStop}%
\bibitem [{\citenamefont {{Shaymatov}}\ \emph {et~al.}(2023)\citenamefont {{Shaymatov}}, \citenamefont {{Ahmedov}}, \citenamefont {{De Laurentis}}, \citenamefont {{Jamil}}, \citenamefont {{Wu}}, \citenamefont {{Wang}},\ and\ \citenamefont {{Azreg-A{\"\i}nou}}}]{Shaymatov23ApJ}%
  \BibitemOpen
  \bibfield  {author} {\bibinfo {author} {\bibfnamefont {S.}~\bibnamefont {{Shaymatov}}}, \bibinfo {author} {\bibfnamefont {B.}~\bibnamefont {{Ahmedov}}}, \bibinfo {author} {\bibfnamefont {M.}~\bibnamefont {{De Laurentis}}}, \bibinfo {author} {\bibfnamefont {M.}~\bibnamefont {{Jamil}}}, \bibinfo {author} {\bibfnamefont {Q.}~\bibnamefont {{Wu}}}, \bibinfo {author} {\bibfnamefont {A.}~\bibnamefont {{Wang}}}, \ and\ \bibinfo {author} {\bibfnamefont {M.}~\bibnamefont {{Azreg-A{\"\i}nou}}},\ }\href {\doibase 10.3847/1538-4357/acfcba} {\bibfield  {journal} {\bibinfo  {journal} {Astrophys. J.}\ }\textbf {\bibinfo {volume} {959}},\ \bibinfo {eid} {6} (\bibinfo {year} {2023})},\ \Eprint {http://arxiv.org/abs/2307.10804} {arXiv:2307.10804 [gr-qc]} \BibitemShut {NoStop}%
\bibitem [{\citenamefont {{Rayimbaev}}\ \emph {et~al.}(2022)\citenamefont {{Rayimbaev}}, \citenamefont {{Kurbonov}}, \citenamefont {{Abdujabbarov}},\ and\ \citenamefont {{Han}}}]{Rayimbaev22IJMPD}%
  \BibitemOpen
  \bibfield  {author} {\bibinfo {author} {\bibfnamefont {J.}~\bibnamefont {{Rayimbaev}}}, \bibinfo {author} {\bibfnamefont {N.}~\bibnamefont {{Kurbonov}}}, \bibinfo {author} {\bibfnamefont {A.}~\bibnamefont {{Abdujabbarov}}}, \ and\ \bibinfo {author} {\bibfnamefont {W.-B.}\ \bibnamefont {{Han}}},\ }\href {\doibase 10.1142/S0218271822500328} {\bibfield  {journal} {\bibinfo  {journal} {Int. J. Mod. Phys. D.}\ }\textbf {\bibinfo {volume} {31}},\ \bibinfo {eid} {2250032-266} (\bibinfo {year} {2022})}\BibitemShut {NoStop}%
\bibitem [{\citenamefont {{Xamidov}}\ \emph {et~al.}(2024)\citenamefont {{Xamidov}}, \citenamefont {{Shaymatov}}, \citenamefont {{Sheoran}},\ and\ \citenamefont {{Ahmedov}}}]{Xamidov24EPJC}%
  \BibitemOpen
  \bibfield  {author} {\bibinfo {author} {\bibfnamefont {T.}~\bibnamefont {{Xamidov}}}, \bibinfo {author} {\bibfnamefont {S.}~\bibnamefont {{Shaymatov}}}, \bibinfo {author} {\bibfnamefont {P.}~\bibnamefont {{Sheoran}}}, \ and\ \bibinfo {author} {\bibfnamefont {B.}~\bibnamefont {{Ahmedov}}},\ }\href {\doibase 10.1140/epjc/s10052-024-13658-w} {\bibfield  {journal} {\bibinfo  {journal} {Eur. Phys. J. C}\ }\textbf {\bibinfo {volume} {84}},\ \bibinfo {eid} {1300} (\bibinfo {year} {2024})},\ \Eprint {http://arxiv.org/abs/2407.19800} {arXiv:2407.19800 [gr-qc]} \BibitemShut {NoStop}%
\bibitem [{\citenamefont {Perlick}\ \emph {et~al.}(2018)\citenamefont {Perlick}, \citenamefont {Tsupko},\ and\ \citenamefont {Bisnovatyi-Kogan}}]{Perlick18}%
  \BibitemOpen
  \bibfield  {author} {\bibinfo {author} {\bibfnamefont {V.}~\bibnamefont {Perlick}}, \bibinfo {author} {\bibfnamefont {O.~Y.}\ \bibnamefont {Tsupko}}, \ and\ \bibinfo {author} {\bibfnamefont {G.~S.}\ \bibnamefont {Bisnovatyi-Kogan}},\ }\href {\doibase 10.1103/PhysRevD.97.104062} {\bibfield  {journal} {\bibinfo  {journal} {Phys. Rev. D}\ }\textbf {\bibinfo {volume} {97}},\ \bibinfo {pages} {104062} (\bibinfo {year} {2018})},\ \Eprint {http://arxiv.org/abs/1804.04898} {arXiv:1804.04898 [gr-qc]} \BibitemShut {NoStop}%
\bibitem [{\citenamefont {Bisnovatyi-Kogan}\ and\ \citenamefont {Tsupko}(2017)}]{Bisnovatyi-Kogan17}%
  \BibitemOpen
  \bibfield  {author} {\bibinfo {author} {\bibfnamefont {G.~S.}\ \bibnamefont {Bisnovatyi-Kogan}}\ and\ \bibinfo {author} {\bibfnamefont {O.~Y.}\ \bibnamefont {Tsupko}},\ }\href {\doibase 10.3390/universe3030057} {\bibfield  {journal} {\bibinfo  {journal} {Universe}\ }\textbf {\bibinfo {volume} {3}},\ \bibinfo {pages} {57} (\bibinfo {year} {2017})},\ \Eprint {http://arxiv.org/abs/1905.06615} {arXiv:1905.06615 [gr-qc]} \BibitemShut {NoStop}%
\bibitem [{\citenamefont {Rogers}(2015)}]{Rogers15MNRAS}%
  \BibitemOpen
  \bibfield  {author} {\bibinfo {author} {\bibfnamefont {A.}~\bibnamefont {Rogers}},\ }\href {\doibase 10.1093/mnras/stv917} {\bibfield  {journal} {\bibinfo  {journal} {Mon. Not. R. Astron. Soc.}\ }\textbf {\bibinfo {volume} {451}},\ \bibinfo {pages} {17} (\bibinfo {year} {2015})},\ \Eprint {http://arxiv.org/abs/1505.06790} {arXiv:1505.06790 [astro-ph.HE]} \BibitemShut {NoStop}%
\bibitem [{\citenamefont {{Atamurotov}}\ \emph {et~al.}(2021)\citenamefont {{Atamurotov}}, \citenamefont {{Shaymatov}}, \citenamefont {{Sheoran}},\ and\ \citenamefont {{Siwach}}}]{Atamurotov21JCAP}%
  \BibitemOpen
  \bibfield  {author} {\bibinfo {author} {\bibfnamefont {F.}~\bibnamefont {{Atamurotov}}}, \bibinfo {author} {\bibfnamefont {S.}~\bibnamefont {{Shaymatov}}}, \bibinfo {author} {\bibfnamefont {P.}~\bibnamefont {{Sheoran}}}, \ and\ \bibinfo {author} {\bibfnamefont {S.}~\bibnamefont {{Siwach}}},\ }\href {\doibase 10.1088/1475-7516/2021/08/045} {\bibfield  {journal} {\bibinfo  {journal} {JCAP}\ }\textbf {\bibinfo {volume} {2021}},\ \bibinfo {eid} {045} (\bibinfo {year} {2021})},\ \Eprint {http://arxiv.org/abs/2105.02214} {arXiv:2105.02214 [gr-qc]} \BibitemShut {NoStop}%
\bibitem [{\citenamefont {Jafarzade}\ \emph {et~al.}(2025)\citenamefont {Jafarzade}, \citenamefont {Shaymatov},\ and\ \citenamefont {Jamil}}]{Jafarzade25APh}%
  \BibitemOpen
  \bibfield  {author} {\bibinfo {author} {\bibfnamefont {K.}~\bibnamefont {Jafarzade}}, \bibinfo {author} {\bibfnamefont {S.}~\bibnamefont {Shaymatov}}, \ and\ \bibinfo {author} {\bibfnamefont {M.}~\bibnamefont {Jamil}},\ }\href {\doibase https://doi.org/10.1016/j.astropartphys.2025.103100} {\bibfield  {journal} {\bibinfo  {journal} {Astropart. Phys.}\ }\textbf {\bibinfo {volume} {168}},\ \bibinfo {pages} {103100} (\bibinfo {year} {2025})},\ \Eprint {http://arxiv.org/abs/2502.14980} {arXiv:2502.14980 [gr-qc]} \BibitemShut {NoStop}%
\bibitem [{\citenamefont {Shaymatov}\ \emph {et~al.}(2022)\citenamefont {Shaymatov}, \citenamefont {Sheoran}, \citenamefont {Becerril}, \citenamefont {Nucamendi},\ and\ \citenamefont {Ahmedov}}]{Shaymatov22PRD}%
  \BibitemOpen
  \bibfield  {author} {\bibinfo {author} {\bibfnamefont {S.}~\bibnamefont {Shaymatov}}, \bibinfo {author} {\bibfnamefont {P.}~\bibnamefont {Sheoran}}, \bibinfo {author} {\bibfnamefont {R.}~\bibnamefont {Becerril}}, \bibinfo {author} {\bibfnamefont {U.}~\bibnamefont {Nucamendi}}, \ and\ \bibinfo {author} {\bibfnamefont {B.}~\bibnamefont {Ahmedov}},\ }\href {\doibase 10.1103/PhysRevD.106.024039} {\bibfield  {journal} {\bibinfo  {journal} {Phys. Rev. D}\ }\textbf {\bibinfo {volume} {106}},\ \bibinfo {pages} {024039} (\bibinfo {year} {2022})},\ \Eprint {http://arxiv.org/abs/2204.02026} {arXiv:2204.02026 [gr-qc]} \BibitemShut {NoStop}%
\bibitem [{\citenamefont {{Xamidov}}\ \emph {et~al.}(2025)\citenamefont {{Xamidov}}, \citenamefont {{Sheoran}}, \citenamefont {{Shaymatov}},\ and\ \citenamefont {{Zhu}}}]{Xamidov25JCAP}%
  \BibitemOpen
  \bibfield  {author} {\bibinfo {author} {\bibfnamefont {T.}~\bibnamefont {{Xamidov}}}, \bibinfo {author} {\bibfnamefont {P.}~\bibnamefont {{Sheoran}}}, \bibinfo {author} {\bibfnamefont {S.}~\bibnamefont {{Shaymatov}}}, \ and\ \bibinfo {author} {\bibfnamefont {T.}~\bibnamefont {{Zhu}}},\ }\href {\doibase 10.1088/1475-7516/2025/03/053} {\bibfield  {journal} {\bibinfo  {journal} {J. Cosmol. Astropart. Phys.}\ }\textbf {\bibinfo {volume} {2025}},\ \bibinfo {eid} {053} (\bibinfo {year} {2025})},\ \Eprint {http://arxiv.org/abs/2412.03885} {arXiv:2412.03885 [gr-qc]} \BibitemShut {NoStop}%
\bibitem [{\citenamefont {{Miao}}\ and\ \citenamefont {{Yang}}(2025)}]{Si-Jiang25JCAP}%
  \BibitemOpen
  \bibfield  {author} {\bibinfo {author} {\bibfnamefont {W.-J.}\ \bibnamefont {{Miao}}}\ and\ \bibinfo {author} {\bibfnamefont {S.-J.}\ \bibnamefont {{Yang}}},\ }\href {\doibase 10.1088/1475-7516/2025/05/022} {\bibfield  {journal} {\bibinfo  {journal} {J. Cosmol. Astropart. Phys.}\ }\textbf {\bibinfo {volume} {2025}},\ \bibinfo {eid} {022} (\bibinfo {year} {2025})},\ \Eprint {http://arxiv.org/abs/2409.07305} {arXiv:2409.07305 [gr-qc]} \BibitemShut {NoStop}%
\bibitem [{\citenamefont {{Schwarzschild}}(1916)}]{1916SPAW.......189S}%
  \BibitemOpen
  \bibfield  {author} {\bibinfo {author} {\bibfnamefont {K.}~\bibnamefont {{Schwarzschild}}},\ }\href@noop {} {\bibfield  {journal} {\bibinfo  {journal} {Sitzungsberichte der K\&ouml;niglich Preussischen Akademie der Wissenschaften}\ ,\ \bibinfo {pages} {189}} (\bibinfo {year} {1916})}\BibitemShut {NoStop}%
\bibitem [{\citenamefont {{Abbott}}\ and\ \citenamefont {et~al. {(Virgo and LIGO Scientific Collaborations)}}(2019)}]{Abbott19aPRX}%
  \BibitemOpen
  \bibfield  {author} {\bibinfo {author} {\bibfnamefont {B.~P.}\ \bibnamefont {{Abbott}}}\ and\ \bibinfo {author} {\bibnamefont {et~al. {(Virgo and LIGO Scientific Collaborations)}}},\ }\href {\doibase 10.1103/PhysRevX.9.031040} {\bibfield  {journal} {\bibinfo  {journal} {Phys. Rev. X}\ }\textbf {\bibinfo {volume} {9}},\ \bibinfo {pages} {031040} (\bibinfo {year} {2019})}\BibitemShut {NoStop}%
\bibitem [{\citenamefont {{Abbott}}\ and\ \citenamefont {et~al. {(Virgo and LIGO Scientific Collaborations)}}(2024)}]{Abbott24:PhysRevD.109.022001}%
  \BibitemOpen
  \bibfield  {author} {\bibinfo {author} {\bibfnamefont {R.}~\bibnamefont {{Abbott}}}\ and\ \bibinfo {author} {\bibnamefont {et~al. {(Virgo and LIGO Scientific Collaborations)}}},\ }\href {\doibase 10.1103/PhysRevD.109.022001} {\bibfield  {journal} {\bibinfo  {journal} {Phys. Rev. D}\ }\textbf {\bibinfo {volume} {109}},\ \bibinfo {pages} {022001} (\bibinfo {year} {2024})}\BibitemShut {NoStop}%
\bibitem [{\citenamefont {{Abbott}}\ and\ \citenamefont {et~al. {(Virgo and LIGO Scientific Collaborations)}}(2023)}]{Abbott23PRX}%
  \BibitemOpen
  \bibfield  {author} {\bibinfo {author} {\bibfnamefont {R.}~\bibnamefont {{Abbott}}}\ and\ \bibinfo {author} {\bibnamefont {et~al. {(Virgo and LIGO Scientific Collaborations)}}},\ }\href {\doibase 10.1103/PhysRevX.13.041039} {\bibfield  {journal} {\bibinfo  {journal} {Phys. Rev. X}\ }\textbf {\bibinfo {volume} {13}},\ \bibinfo {pages} {041039} (\bibinfo {year} {2023})}\BibitemShut {NoStop}%
\bibitem [{\citenamefont {{Amaro-Seoane}}\ and\ \citenamefont {et~al. {(Laser Interferometer Space Antenna)}}(2017)}]{Amaro-Seoane2017LISA}%
  \BibitemOpen
  \bibfield  {author} {\bibinfo {author} {\bibfnamefont {P.}~\bibnamefont {{Amaro-Seoane}}}\ and\ \bibinfo {author} {\bibnamefont {et~al. {(Laser Interferometer Space Antenna)}}},\ }\href {https://arxiv.org/abs/1702.00786} {\enquote {\bibinfo {title} {Laser interferometer space antenna},}\ } (\bibinfo {year} {2017}),\ \Eprint {http://arxiv.org/abs/1702.00786} {arXiv:1702.00786 [astro-ph.IM]} \BibitemShut {NoStop}%
\bibitem [{\citenamefont {Hu}\ and\ \citenamefont {Wu}(2017)}]{Hu:10.1093/nsr/nwx116}%
  \BibitemOpen
  \bibfield  {author} {\bibinfo {author} {\bibfnamefont {W.-R.}\ \bibnamefont {Hu}}\ and\ \bibinfo {author} {\bibfnamefont {Y.-L.}\ \bibnamefont {Wu}},\ }\href {\doibase 10.1093/nsr/nwx116} {\bibfield  {journal} {\bibinfo  {journal} {Natl. Sci. Rev.}\ }\textbf {\bibinfo {volume} {4}},\ \bibinfo {pages} {685} (\bibinfo {year} {2017})},\ \Eprint {http://arxiv.org/abs/https://academic.oup.com/nsr/article-pdf/4/5/685/31566708/nwx116.pdf} {https://academic.oup.com/nsr/article-pdf/4/5/685/31566708/nwx116.pdf} \BibitemShut {NoStop}%
\bibitem [{\citenamefont {{Luo}}\ and\ \citenamefont {et~al.}(2016)}]{Luo:TianQin2016}%
  \BibitemOpen
  \bibfield  {author} {\bibinfo {author} {\bibfnamefont {J.}~\bibnamefont {{Luo}}}\ and\ \bibinfo {author} {\bibnamefont {et~al.}},\ }\href {\doibase 10.1088/0264-9381/33/3/035010} {\bibfield  {journal} {\bibinfo  {journal} {Class. Quantum Grav.}\ }\textbf {\bibinfo {volume} {33}},\ \bibinfo {eid} {035010} (\bibinfo {year} {2016})},\ \Eprint {http://arxiv.org/abs/1512.02076} {arXiv:1512.02076 [astro-ph.IM]} \BibitemShut {NoStop}%
\bibitem [{\citenamefont {{Baibhav}}\ and\ \citenamefont {et~al.}(2021)}]{Baibhav21ExA}%
  \BibitemOpen
  \bibfield  {author} {\bibinfo {author} {\bibfnamefont {V.}~\bibnamefont {{Baibhav}}}\ and\ \bibinfo {author} {\bibnamefont {et~al.}},\ }\href {\doibase 10.1007/s10686-021-09741-9} {\bibfield  {journal} {\bibinfo  {journal} {Exp. Astron.}\ }\textbf {\bibinfo {volume} {51}},\ \bibinfo {pages} {1385} (\bibinfo {year} {2021})},\ \Eprint {http://arxiv.org/abs/1908.11390} {arXiv:1908.11390 [astro-ph.HE]} \BibitemShut {NoStop}%
\bibitem [{\citenamefont {{Amaro-Seoane}}\ and\ \citenamefont {et~al.}(2023)}]{Amaro-Seoane2023LRR}%
  \BibitemOpen
  \bibfield  {author} {\bibinfo {author} {\bibfnamefont {P.}~\bibnamefont {{Amaro-Seoane}}}\ and\ \bibinfo {author} {\bibnamefont {et~al.}},\ }\href {\doibase 10.1007/s41114-022-00041-y} {\bibfield  {journal} {\bibinfo  {journal} {Living Rev. Relativ.}\ }\textbf {\bibinfo {volume} {26}},\ \bibinfo {eid} {2} (\bibinfo {year} {2023})},\ \Eprint {http://arxiv.org/abs/2203.06016} {arXiv:2203.06016 [gr-qc]} \BibitemShut {NoStop}%
\bibitem [{\citenamefont {{Arun}}\ and\ \citenamefont {et~al. {(LISA Collaboration)}}(2022)}]{Arun22LRR:LISA}%
  \BibitemOpen
  \bibfield  {author} {\bibinfo {author} {\bibfnamefont {K.~G.}\ \bibnamefont {{Arun}}}\ and\ \bibinfo {author} {\bibnamefont {et~al. {(LISA Collaboration)}}},\ }\href {\doibase 10.1007/s41114-022-00036-9} {\bibfield  {journal} {\bibinfo  {journal} {Living Rev. Relativ.}\ }\textbf {\bibinfo {volume} {25}},\ \bibinfo {eid} {4} (\bibinfo {year} {2022})},\ \Eprint {http://arxiv.org/abs/2205.01597} {arXiv:2205.01597 [gr-qc]} \BibitemShut {NoStop}%
\bibitem [{\citenamefont {Hughes}(2001)}]{Hughes2001CQG}%
  \BibitemOpen
  \bibfield  {author} {\bibinfo {author} {\bibfnamefont {S.~A.}\ \bibnamefont {Hughes}},\ }\href {\doibase 10.1088/0264-9381/18/19/314} {\bibfield  {journal} {\bibinfo  {journal} {Class. Quantum Gravity}\ }\textbf {\bibinfo {volume} {18}},\ \bibinfo {pages} {4067–4073} (\bibinfo {year} {2001})}\BibitemShut {NoStop}%
\bibitem [{\citenamefont {{Amaro-Seoane}}(2018)}]{Amaro-Seoane18LRR}%
  \BibitemOpen
  \bibfield  {author} {\bibinfo {author} {\bibfnamefont {P.}~\bibnamefont {{Amaro-Seoane}}},\ }\href {\doibase 10.1007/s41114-018-0013-8} {\bibfield  {journal} {\bibinfo  {journal} {Living Rev. Rel.}\ }\textbf {\bibinfo {volume} {21}},\ \bibinfo {eid} {4} (\bibinfo {year} {2018})},\ \Eprint {http://arxiv.org/abs/1205.5240} {arXiv:1205.5240 [astro-ph.CO]} \BibitemShut {NoStop}%
\bibitem [{\citenamefont {{Babak}}\ \emph {et~al.}(2017)\citenamefont {{Babak}}, \citenamefont {{Gair}}, \citenamefont {{Sesana}}, \citenamefont {{Barausse}}, \citenamefont {{Sopuerta}}, \citenamefont {{Berry}}, \citenamefont {{Berti}}, \citenamefont {{Amaro-Seoane}}, \citenamefont {{Petiteau}},\ and\ \citenamefont {{Klein}}}]{Babak17PRD}%
  \BibitemOpen
  \bibfield  {author} {\bibinfo {author} {\bibfnamefont {S.}~\bibnamefont {{Babak}}}, \bibinfo {author} {\bibfnamefont {J.}~\bibnamefont {{Gair}}}, \bibinfo {author} {\bibfnamefont {A.}~\bibnamefont {{Sesana}}}, \bibinfo {author} {\bibfnamefont {E.}~\bibnamefont {{Barausse}}}, \bibinfo {author} {\bibfnamefont {C.~F.}\ \bibnamefont {{Sopuerta}}}, \bibinfo {author} {\bibfnamefont {C.~P.~L.}\ \bibnamefont {{Berry}}}, \bibinfo {author} {\bibfnamefont {E.}~\bibnamefont {{Berti}}}, \bibinfo {author} {\bibfnamefont {P.}~\bibnamefont {{Amaro-Seoane}}}, \bibinfo {author} {\bibfnamefont {A.}~\bibnamefont {{Petiteau}}}, \ and\ \bibinfo {author} {\bibfnamefont {A.}~\bibnamefont {{Klein}}},\ }\href {\doibase 10.1103/PhysRevD.95.103012} {\bibfield  {journal} {\bibinfo  {journal} {Phys. Rev. D}\ }\textbf {\bibinfo {volume} {95}},\ \bibinfo {eid} {103012} (\bibinfo {year} {2017})},\ \Eprint {http://arxiv.org/abs/1703.09722} {arXiv:1703.09722 [gr-qc]} \BibitemShut {NoStop}%
\bibitem [{\citenamefont {{Amaro-Seoane}}\ \emph {et~al.}(2007)\citenamefont {{Amaro-Seoane}}, \citenamefont {{Gair}}, \citenamefont {{Freitag}}, \citenamefont {{Miller}}, \citenamefont {{Mandel}}, \citenamefont {{Cutler}},\ and\ \citenamefont {{Babak}}}]{Amaro-Seoane2007CQG}%
  \BibitemOpen
  \bibfield  {author} {\bibinfo {author} {\bibfnamefont {P.}~\bibnamefont {{Amaro-Seoane}}}, \bibinfo {author} {\bibfnamefont {J.~R.}\ \bibnamefont {{Gair}}}, \bibinfo {author} {\bibfnamefont {M.}~\bibnamefont {{Freitag}}}, \bibinfo {author} {\bibfnamefont {M.~C.}\ \bibnamefont {{Miller}}}, \bibinfo {author} {\bibfnamefont {I.}~\bibnamefont {{Mandel}}}, \bibinfo {author} {\bibfnamefont {C.~J.}\ \bibnamefont {{Cutler}}}, \ and\ \bibinfo {author} {\bibfnamefont {S.}~\bibnamefont {{Babak}}},\ }\href {\doibase 10.1088/0264-9381/24/17/R01} {\bibfield  {journal} {\bibinfo  {journal} {Class. Quantum Grav.}\ }\textbf {\bibinfo {volume} {24}},\ \bibinfo {pages} {R113} (\bibinfo {year} {2007})},\ \Eprint {http://arxiv.org/abs/astro-ph/0703495} {arXiv:astro-ph/0703495 [astro-ph]} \BibitemShut {NoStop}%
\bibitem [{\citenamefont {{McGee}}\ \emph {et~al.}(2020)\citenamefont {{McGee}}, \citenamefont {{Sesana}},\ and\ \citenamefont {{Vecchio}}}]{McGee2020NatAs}%
  \BibitemOpen
  \bibfield  {author} {\bibinfo {author} {\bibfnamefont {S.}~\bibnamefont {{McGee}}}, \bibinfo {author} {\bibfnamefont {A.}~\bibnamefont {{Sesana}}}, \ and\ \bibinfo {author} {\bibfnamefont {A.}~\bibnamefont {{Vecchio}}},\ }\href {\doibase 10.1038/s41550-019-0969-7} {\bibfield  {journal} {\bibinfo  {journal} {Nat. Astron.}\ }\textbf {\bibinfo {volume} {4}},\ \bibinfo {pages} {26} (\bibinfo {year} {2020})},\ \Eprint {http://arxiv.org/abs/1811.00050} {arXiv:1811.00050 [astro-ph.HE]} \BibitemShut {NoStop}%
\bibitem [{\citenamefont {{Laghi}}\ \emph {et~al.}(2021)\citenamefont {{Laghi}}, \citenamefont {{Tamanini}}, \citenamefont {{Del Pozzo}}, \citenamefont {{Sesana}}, \citenamefont {{Gair}}, \citenamefont {{Babak}},\ and\ \citenamefont {{Izquierdo-Villalba}}}]{Laghi2021MNRAS}%
  \BibitemOpen
  \bibfield  {author} {\bibinfo {author} {\bibfnamefont {D.}~\bibnamefont {{Laghi}}}, \bibinfo {author} {\bibfnamefont {N.}~\bibnamefont {{Tamanini}}}, \bibinfo {author} {\bibfnamefont {W.}~\bibnamefont {{Del Pozzo}}}, \bibinfo {author} {\bibfnamefont {A.}~\bibnamefont {{Sesana}}}, \bibinfo {author} {\bibfnamefont {J.}~\bibnamefont {{Gair}}}, \bibinfo {author} {\bibfnamefont {S.}~\bibnamefont {{Babak}}}, \ and\ \bibinfo {author} {\bibfnamefont {D.}~\bibnamefont {{Izquierdo-Villalba}}},\ }\href {\doibase 10.1093/mnras/stab2741} {\bibfield  {journal} {\bibinfo  {journal} {Mon. Not. R. Astron. Soc.}\ }\textbf {\bibinfo {volume} {508}},\ \bibinfo {pages} {4512} (\bibinfo {year} {2021})},\ \Eprint {http://arxiv.org/abs/2102.01708} {arXiv:2102.01708 [astro-ph.CO]} \BibitemShut {NoStop}%
\bibitem [{\citenamefont {{Amaro Seoane}}\ and\ \citenamefont {et~al.}(2022)}]{Amaro-Seoane2022GRG}%
  \BibitemOpen
  \bibfield  {author} {\bibinfo {author} {\bibfnamefont {P.}~\bibnamefont {{Amaro Seoane}}}\ and\ \bibinfo {author} {\bibnamefont {et~al.}},\ }\href {\doibase 10.1007/s10714-021-02889-x} {\bibfield  {journal} {\bibinfo  {journal} {Gen. Relativ. Gravit.}\ }\textbf {\bibinfo {volume} {54}},\ \bibinfo {eid} {3} (\bibinfo {year} {2022})},\ \Eprint {http://arxiv.org/abs/2107.09665} {arXiv:2107.09665 [astro-ph.IM]} \BibitemShut {NoStop}%
\bibitem [{\citenamefont {Levin}\ and\ \citenamefont {Perez-Giz}(2008)}]{Levin_2008}%
  \BibitemOpen
  \bibfield  {author} {\bibinfo {author} {\bibfnamefont {J.}~\bibnamefont {Levin}}\ and\ \bibinfo {author} {\bibfnamefont {G.}~\bibnamefont {Perez-Giz}},\ }\href {\doibase 10.1103/physrevd.77.103005} {\bibfield  {journal} {\bibinfo  {journal} {Phys. Rev. D}\ }\textbf {\bibinfo {volume} {77}} (\bibinfo {year} {2008}),\ 10.1103/physrevd.77.103005}\BibitemShut {NoStop}%
\bibitem [{\citenamefont {Grossman}\ and\ \citenamefont {Levin}(2009)}]{Grossman_2009}%
  \BibitemOpen
  \bibfield  {author} {\bibinfo {author} {\bibfnamefont {R.}~\bibnamefont {Grossman}}\ and\ \bibinfo {author} {\bibfnamefont {J.}~\bibnamefont {Levin}},\ }\href {\doibase 10.1103/physrevd.79.043017} {\bibfield  {journal} {\bibinfo  {journal} {Phys. Rev. D}\ }\textbf {\bibinfo {volume} {79}} (\bibinfo {year} {2009}),\ 10.1103/physrevd.79.043017}\BibitemShut {NoStop}%
\bibitem [{\citenamefont {Misra}\ and\ \citenamefont {Levin}(2010{\natexlab{a}})}]{Misra_2010}%
  \BibitemOpen
  \bibfield  {author} {\bibinfo {author} {\bibfnamefont {V.}~\bibnamefont {Misra}}\ and\ \bibinfo {author} {\bibfnamefont {J.}~\bibnamefont {Levin}},\ }\href {\doibase 10.1103/physrevd.82.083001} {\bibfield  {journal} {\bibinfo  {journal} {Phys. Rev. D}\ }\textbf {\bibinfo {volume} {82}} (\bibinfo {year} {2010}{\natexlab{a}}),\ 10.1103/physrevd.82.083001}\BibitemShut {NoStop}%
\bibitem [{\citenamefont {Levin}(2009)}]{Levin_2009}%
  \BibitemOpen
  \bibfield  {author} {\bibinfo {author} {\bibfnamefont {J.}~\bibnamefont {Levin}},\ }\href {\doibase 10.1088/0264-9381/26/23/235010} {\bibfield  {journal} {\bibinfo  {journal} {Class. Quantum Grav.}\ }\textbf {\bibinfo {volume} {26}},\ \bibinfo {pages} {235010} (\bibinfo {year} {2009})}\BibitemShut {NoStop}%
\bibitem [{\citenamefont {{Glampedakis}}\ and\ \citenamefont {{Kennefick}}(2002)}]{Glampedakis02PRD}%
  \BibitemOpen
  \bibfield  {author} {\bibinfo {author} {\bibfnamefont {K.}~\bibnamefont {{Glampedakis}}}\ and\ \bibinfo {author} {\bibfnamefont {D.}~\bibnamefont {{Kennefick}}},\ }\href {\doibase 10.1103/PhysRevD.66.044002} {\bibfield  {journal} {\bibinfo  {journal} {Phys. Rev. D}\ }\textbf {\bibinfo {volume} {66}},\ \bibinfo {eid} {044002} (\bibinfo {year} {2002})},\ \Eprint {http://arxiv.org/abs/gr-qc/0203086} {arXiv:gr-qc/0203086 [gr-qc]} \BibitemShut {NoStop}%
\bibitem [{\citenamefont {Misra}\ and\ \citenamefont {Levin}(2010{\natexlab{b}})}]{Levin2010}%
  \BibitemOpen
  \bibfield  {author} {\bibinfo {author} {\bibfnamefont {V.}~\bibnamefont {Misra}}\ and\ \bibinfo {author} {\bibfnamefont {J.}~\bibnamefont {Levin}},\ }\href {\doibase 10.1103/PhysRevD.82.083001} {\bibfield  {journal} {\bibinfo  {journal} {Phys. Rev. D}\ }\textbf {\bibinfo {volume} {82}},\ \bibinfo {pages} {083001} (\bibinfo {year} {2010}{\natexlab{b}})}\BibitemShut {NoStop}%
\bibitem [{\citenamefont {{Babar}}\ \emph {et~al.}(2017)\citenamefont {{Babar}}, \citenamefont {{Babar}},\ and\ \citenamefont {{Lim}}}]{Babar17PRD}%
  \BibitemOpen
  \bibfield  {author} {\bibinfo {author} {\bibfnamefont {G.~Z.}\ \bibnamefont {{Babar}}}, \bibinfo {author} {\bibfnamefont {A.~Z.}\ \bibnamefont {{Babar}}}, \ and\ \bibinfo {author} {\bibfnamefont {Y.-K.}\ \bibnamefont {{Lim}}},\ }\href {\doibase 10.1103/PhysRevD.96.084052} {\bibfield  {journal} {\bibinfo  {journal} {Phys. Rev. D}\ }\textbf {\bibinfo {volume} {96}},\ \bibinfo {eid} {084052} (\bibinfo {year} {2017})},\ \Eprint {http://arxiv.org/abs/1710.09581} {arXiv:1710.09581 [gr-qc]} \BibitemShut {NoStop}%
\bibitem [{\citenamefont {Liu}\ \emph {et~al.}(2019)\citenamefont {Liu}, \citenamefont {Ding},\ and\ \citenamefont {Jing}}]{Liu_2019}%
  \BibitemOpen
  \bibfield  {author} {\bibinfo {author} {\bibfnamefont {C.-Q.}\ \bibnamefont {Liu}}, \bibinfo {author} {\bibfnamefont {C.-K.}\ \bibnamefont {Ding}}, \ and\ \bibinfo {author} {\bibfnamefont {J.-L.}\ \bibnamefont {Jing}},\ }\href {\doibase 10.1088/0253-6102/71/12/1461} {\bibfield  {journal} {\bibinfo  {journal} {Commun. Theor. Phys.}\ }\textbf {\bibinfo {volume} {71}},\ \bibinfo {pages} {1461} (\bibinfo {year} {2019})}\BibitemShut {NoStop}%
\bibitem [{\citenamefont {{Deng}}(2020)}]{Deng20}%
  \BibitemOpen
  \bibfield  {author} {\bibinfo {author} {\bibfnamefont {X.-M.}\ \bibnamefont {{Deng}}},\ }\href {\doibase 10.1016/j.dark.2020.100629} {\bibfield  {journal} {\bibinfo  {journal} {Phys. Dark Universe}\ }\textbf {\bibinfo {volume} {30}},\ \bibinfo {eid} {100629} (\bibinfo {year} {2020})}\BibitemShut {NoStop}%
\bibitem [{\citenamefont {Wei}\ \emph {et~al.}(2019)\citenamefont {Wei}, \citenamefont {Yang},\ and\ \citenamefont {Liu}}]{Wei2019PRD}%
  \BibitemOpen
  \bibfield  {author} {\bibinfo {author} {\bibfnamefont {S.-W.}\ \bibnamefont {Wei}}, \bibinfo {author} {\bibfnamefont {J.}~\bibnamefont {Yang}}, \ and\ \bibinfo {author} {\bibfnamefont {Y.-X.}\ \bibnamefont {Liu}},\ }\href {\doibase 10.1103/PhysRevD.99.104016} {\bibfield  {journal} {\bibinfo  {journal} {Phys. Rev. D}\ }\textbf {\bibinfo {volume} {99}},\ \bibinfo {pages} {104016} (\bibinfo {year} {2019})}\BibitemShut {NoStop}%
\bibitem [{\citenamefont {{Jiang}}\ \emph {et~al.}(2024)\citenamefont {{Jiang}}, \citenamefont {{Alloqulov}}, \citenamefont {{Wu}}, \citenamefont {{Shaymatov}},\ and\ \citenamefont {{Zhu}}}]{Jiang2024PDU}%
  \BibitemOpen
  \bibfield  {author} {\bibinfo {author} {\bibfnamefont {H.}~\bibnamefont {{Jiang}}}, \bibinfo {author} {\bibfnamefont {M.}~\bibnamefont {{Alloqulov}}}, \bibinfo {author} {\bibfnamefont {Q.}~\bibnamefont {{Wu}}}, \bibinfo {author} {\bibfnamefont {S.}~\bibnamefont {{Shaymatov}}}, \ and\ \bibinfo {author} {\bibfnamefont {T.}~\bibnamefont {{Zhu}}},\ }\href {\doibase 10.1016/j.dark.2024.101627} {\bibfield  {journal} {\bibinfo  {journal} {Phys. Dark Universe}\ }\textbf {\bibinfo {volume} {46}},\ \bibinfo {eid} {101627} (\bibinfo {year} {2024})}\BibitemShut {NoStop}%
\bibitem [{\citenamefont {Tu}\ \emph {et~al.}(2023)\citenamefont {Tu}, \citenamefont {Zhu},\ and\ \citenamefont {Wang}}]{Tu23PRD}%
  \BibitemOpen
  \bibfield  {author} {\bibinfo {author} {\bibfnamefont {Z.-Y.}\ \bibnamefont {Tu}}, \bibinfo {author} {\bibfnamefont {T.}~\bibnamefont {Zhu}}, \ and\ \bibinfo {author} {\bibfnamefont {A.}~\bibnamefont {Wang}},\ }\href {\doibase 10.1103/PhysRevD.108.024035} {\bibfield  {journal} {\bibinfo  {journal} {Phys. Rev. D}\ }\textbf {\bibinfo {volume} {108}},\ \bibinfo {pages} {024035} (\bibinfo {year} {2023})}\BibitemShut {NoStop}%
\bibitem [{\citenamefont {Chen}\ and\ \citenamefont {Yang}(2025)}]{Chen:2025aqh}%
  \BibitemOpen
  \bibfield  {author} {\bibinfo {author} {\bibfnamefont {J.}~\bibnamefont {Chen}}\ and\ \bibinfo {author} {\bibfnamefont {J.}~\bibnamefont {Yang}},\ }\href {\doibase 10.1140/epjc/s10052-025-14457-7} {\bibfield  {journal} {\bibinfo  {journal} {Eur. Phys. J. C}\ }\textbf {\bibinfo {volume} {85}},\ \bibinfo {pages} {726} (\bibinfo {year} {2025})},\ \Eprint {http://arxiv.org/abs/2505.02660} {arXiv:2505.02660 [gr-qc]} \BibitemShut {NoStop}%
\bibitem [{\citenamefont {Lu}\ and\ \citenamefont {Zhu}(2025)}]{Lu:2025cxx}%
  \BibitemOpen
  \bibfield  {author} {\bibinfo {author} {\bibfnamefont {S.}~\bibnamefont {Lu}}\ and\ \bibinfo {author} {\bibfnamefont {T.}~\bibnamefont {Zhu}},\ }\href@noop {} {\  (\bibinfo {year} {2025})},\ \Eprint {http://arxiv.org/abs/2505.00294} {arXiv:2505.00294 [gr-qc]} \BibitemShut {NoStop}%
\bibitem [{\citenamefont {Meng}\ \emph {et~al.}(2025)\citenamefont {Meng}, \citenamefont {Xu},\ and\ \citenamefont {Tang}}]{meng2025periodicorbitsgravitationalwave}%
  \BibitemOpen
  \bibfield  {author} {\bibinfo {author} {\bibfnamefont {L.}~\bibnamefont {Meng}}, \bibinfo {author} {\bibfnamefont {Z.}~\bibnamefont {Xu}}, \ and\ \bibinfo {author} {\bibfnamefont {M.}~\bibnamefont {Tang}},\ }\href {https://arxiv.org/abs/2506.05015} {\enquote {\bibinfo {title} {Periodic orbits and gravitational wave radiation of black holes in 4d-egb gravity},}\ } (\bibinfo {year} {2025}),\ \Eprint {http://arxiv.org/abs/2506.05015} {arXiv:2506.05015 [gr-qc]} \BibitemShut {NoStop}%
\bibitem [{\citenamefont {{Barausse}}\ \emph {et~al.}(2014)\citenamefont {{Barausse}}, \citenamefont {{Cardoso}},\ and\ \citenamefont {{Pani}}}]{Barausse14PRD}%
  \BibitemOpen
  \bibfield  {author} {\bibinfo {author} {\bibfnamefont {E.}~\bibnamefont {{Barausse}}}, \bibinfo {author} {\bibfnamefont {V.}~\bibnamefont {{Cardoso}}}, \ and\ \bibinfo {author} {\bibfnamefont {P.}~\bibnamefont {{Pani}}},\ }\href {\doibase 10.1103/PhysRevD.89.104059} {\bibfield  {journal} {\bibinfo  {journal} {Phys. Rev. D}\ }\textbf {\bibinfo {volume} {89}},\ \bibinfo {eid} {104059} (\bibinfo {year} {2014})},\ \Eprint {http://arxiv.org/abs/1404.7149} {arXiv:1404.7149 [gr-qc]} \BibitemShut {NoStop}%
\bibitem [{\citenamefont {{Cardoso}}\ \emph {et~al.}(2022)\citenamefont {{Cardoso}}, \citenamefont {{Destounis}}, \citenamefont {{Duque}}, \citenamefont {{Macedo}},\ and\ \citenamefont {{Maselli}}}]{Cardoso22PRD}%
  \BibitemOpen
  \bibfield  {author} {\bibinfo {author} {\bibfnamefont {V.}~\bibnamefont {{Cardoso}}}, \bibinfo {author} {\bibfnamefont {K.}~\bibnamefont {{Destounis}}}, \bibinfo {author} {\bibfnamefont {F.}~\bibnamefont {{Duque}}}, \bibinfo {author} {\bibfnamefont {R.~P.}\ \bibnamefont {{Macedo}}}, \ and\ \bibinfo {author} {\bibfnamefont {A.}~\bibnamefont {{Maselli}}},\ }\href {\doibase 10.1103/PhysRevD.105.L061501} {\bibfield  {journal} {\bibinfo  {journal} {Phys. Rev. D}\ }\textbf {\bibinfo {volume} {105}},\ \bibinfo {eid} {L061501} (\bibinfo {year} {2022})},\ \Eprint {http://arxiv.org/abs/2109.00005} {arXiv:2109.00005 [gr-qc]} \BibitemShut {NoStop}%
\bibitem [{\citenamefont {{Yang}}\ \emph {et~al.}(2025{\natexlab{a}})\citenamefont {{Yang}}, \citenamefont {{Zhang}}, \citenamefont {{Zhu}}, \citenamefont {{Zhao}},\ and\ \citenamefont {{Liu}}}]{Yang2025JCAP}%
  \BibitemOpen
  \bibfield  {author} {\bibinfo {author} {\bibfnamefont {S.}~\bibnamefont {{Yang}}}, \bibinfo {author} {\bibfnamefont {Y.-P.}\ \bibnamefont {{Zhang}}}, \bibinfo {author} {\bibfnamefont {T.}~\bibnamefont {{Zhu}}}, \bibinfo {author} {\bibfnamefont {L.}~\bibnamefont {{Zhao}}}, \ and\ \bibinfo {author} {\bibfnamefont {Y.-X.}\ \bibnamefont {{Liu}}},\ }\href {\doibase 10.1088/1475-7516/2025/01/091} {\bibfield  {journal} {\bibinfo  {journal} {J. Cosmol. Astropart. Phys.}\ }\textbf {\bibinfo {volume} {2025}},\ \bibinfo {eid} {091} (\bibinfo {year} {2025}{\natexlab{a}})},\ \Eprint {http://arxiv.org/abs/2407.00283} {arXiv:2407.00283 [gr-qc]} \BibitemShut {NoStop}%
\bibitem [{\citenamefont {Haroon}\ and\ \citenamefont {Zhu}(2025)}]{Haroon2025}%
  \BibitemOpen
  \bibfield  {author} {\bibinfo {author} {\bibfnamefont {S.}~\bibnamefont {Haroon}}\ and\ \bibinfo {author} {\bibfnamefont {T.}~\bibnamefont {Zhu}},\ }\href {https://arxiv.org/abs/2502.09171} {\enquote {\bibinfo {title} {Periodic orbits and their gravitational wave radiations in black hole with dark matter halo},}\ } (\bibinfo {year} {2025}),\ \Eprint {http://arxiv.org/abs/2502.09171} {arXiv:2502.09171 [gr-qc]} \BibitemShut {NoStop}%
\bibitem [{\citenamefont {{Alloqulov}}\ \emph {et~al.}(2025)\citenamefont {{Alloqulov}}, \citenamefont {{Xamidov}}, \citenamefont {{Shaymatov}},\ and\ \citenamefont {{Ahmedov}}}]{Alloqulov25GW1}%
  \BibitemOpen
  \bibfield  {author} {\bibinfo {author} {\bibfnamefont {M.}~\bibnamefont {{Alloqulov}}}, \bibinfo {author} {\bibfnamefont {T.}~\bibnamefont {{Xamidov}}}, \bibinfo {author} {\bibfnamefont {S.}~\bibnamefont {{Shaymatov}}}, \ and\ \bibinfo {author} {\bibfnamefont {B.}~\bibnamefont {{Ahmedov}}},\ }\href {\doibase 10.1140/epjc/s10052-025-14529-8} {\bibfield  {journal} {\bibinfo  {journal} {Eur. Phys. J. C}\ }\textbf {\bibinfo {volume} {85}},\ \bibinfo {eid} {798} (\bibinfo {year} {2025})}\BibitemShut {NoStop}%
\bibitem [{\citenamefont {Ali}\ \emph {et~al.}(2020)\citenamefont {Ali}, \citenamefont {Ghosh},\ and\ \citenamefont {Maharaj}}]{Ali_2020}%
  \BibitemOpen
  \bibfield  {author} {\bibinfo {author} {\bibfnamefont {M.~S.}\ \bibnamefont {Ali}}, \bibinfo {author} {\bibfnamefont {S.~G.}\ \bibnamefont {Ghosh}}, \ and\ \bibinfo {author} {\bibfnamefont {S.~D.}\ \bibnamefont {Maharaj}},\ }\href {\doibase 10.1088/1361-6382/ab9c6c} {\bibfield  {journal} {\bibinfo  {journal} {Classical and Quantum Gravity}\ }\textbf {\bibinfo {volume} {37}},\ \bibinfo {pages} {185003} (\bibinfo {year} {2020})}\BibitemShut {NoStop}%
\bibitem [{\citenamefont {{Babak}}\ \emph {et~al.}(2007)\citenamefont {{Babak}}, \citenamefont {{Fang}}, \citenamefont {{Gair}}, \citenamefont {{Glampedakis}},\ and\ \citenamefont {{Hughes}}}]{Babak07PRD}%
  \BibitemOpen
  \bibfield  {author} {\bibinfo {author} {\bibfnamefont {S.}~\bibnamefont {{Babak}}}, \bibinfo {author} {\bibfnamefont {H.}~\bibnamefont {{Fang}}}, \bibinfo {author} {\bibfnamefont {J.~R.}\ \bibnamefont {{Gair}}}, \bibinfo {author} {\bibfnamefont {K.}~\bibnamefont {{Glampedakis}}}, \ and\ \bibinfo {author} {\bibfnamefont {S.~A.}\ \bibnamefont {{Hughes}}},\ }\href {\doibase 10.1103/PhysRevD.75.024005} {\bibfield  {journal} {\bibinfo  {journal} {Phys. Rev. D}\ }\textbf {\bibinfo {volume} {75}},\ \bibinfo {eid} {024005} (\bibinfo {year} {2007})},\ \Eprint {http://arxiv.org/abs/gr-qc/0607007} {arXiv:gr-qc/0607007 [gr-qc]} \BibitemShut {NoStop}%
\bibitem [{\citenamefont {Chandrasekhar}(1984)}]{1983mtbh.book.....C}%
  \BibitemOpen
  \bibfield  {author} {\bibinfo {author} {\bibfnamefont {S.}~\bibnamefont {Chandrasekhar}},\ }\href {https://doi.org/10.1007/978-94-009-6469-3_2} {\emph {\bibinfo {title} {General Relativity and Gravitation: Invited Papers and Discussion Reports of the 10th International Conference on General Relativity and Gravitation, Padua, July 3--8, 1983}}},\ edited by\ \bibinfo {editor} {\bibfnamefont {B.}~\bibnamefont {Bertotti}}, \bibinfo {editor} {\bibfnamefont {F.}~\bibnamefont {de~Felice}}, \ and\ \bibinfo {editor} {\bibfnamefont {A.}~\bibnamefont {Pascolini}}\ (\bibinfo  {publisher} {Springer Netherlands},\ \bibinfo {address} {Dordrecht},\ \bibinfo {year} {1984})\ pp.\ \bibinfo {pages} {5--26}\BibitemShut {NoStop}%
\bibitem [{\citenamefont {{Yang}}\ \emph {et~al.}(2025{\natexlab{b}})\citenamefont {{Yang}}, \citenamefont {{Zhang}}, \citenamefont {{Zhu}}, \citenamefont {{Zhao}},\ and\ \citenamefont {{Liu}}}]{2025JCAP...01..091Y}%
  \BibitemOpen
  \bibfield  {author} {\bibinfo {author} {\bibfnamefont {S.}~\bibnamefont {{Yang}}}, \bibinfo {author} {\bibfnamefont {Y.-P.}\ \bibnamefont {{Zhang}}}, \bibinfo {author} {\bibfnamefont {T.}~\bibnamefont {{Zhu}}}, \bibinfo {author} {\bibfnamefont {L.}~\bibnamefont {{Zhao}}}, \ and\ \bibinfo {author} {\bibfnamefont {Y.-X.}\ \bibnamefont {{Liu}}},\ }\href {\doibase 10.1088/1475-7516/2025/01/091} {\bibfield  {journal} {\bibinfo  {journal} {Journal of Cosmology and Astroparticle Physics}\ }\textbf {\bibinfo {volume} {2025}},\ \bibinfo {eid} {091} (\bibinfo {year} {2025}{\natexlab{b}})},\ \Eprint {http://arxiv.org/abs/2407.00283} {arXiv:2407.00283 [gr-qc]} \BibitemShut {NoStop}%
\bibitem [{\citenamefont {Shabbir}\ \emph {et~al.}(2025)\citenamefont {Shabbir}, \citenamefont {Jamil},\ and\ \citenamefont {Azreg-Aïnou}}]{SHABBIR2025101816}%
  \BibitemOpen
  \bibfield  {author} {\bibinfo {author} {\bibfnamefont {O.}~\bibnamefont {Shabbir}}, \bibinfo {author} {\bibfnamefont {M.}~\bibnamefont {Jamil}}, \ and\ \bibinfo {author} {\bibfnamefont {M.}~\bibnamefont {Azreg-Aïnou}},\ }\href {\doibase https://doi.org/10.1016/j.dark.2025.101816} {\bibfield  {journal} {\bibinfo  {journal} {Physics of the Dark Universe}\ }\textbf {\bibinfo {volume} {47}},\ \bibinfo {pages} {101816} (\bibinfo {year} {2025})}\BibitemShut {NoStop}%
\bibitem [{\citenamefont {Thorne}(1980)}]{RevModPhys.52.299}%
  \BibitemOpen
  \bibfield  {author} {\bibinfo {author} {\bibfnamefont {K.~S.}\ \bibnamefont {Thorne}},\ }\href {\doibase 10.1103/RevModPhys.52.299} {\bibfield  {journal} {\bibinfo  {journal} {Rev. Mod. Phys.}\ }\textbf {\bibinfo {volume} {52}},\ \bibinfo {pages} {299} (\bibinfo {year} {1980})}\BibitemShut {NoStop}%
\bibitem [{\citenamefont {Poisson}\ and\ \citenamefont {Will}(2014)}]{Poisson_Will_2014}%
  \BibitemOpen
  \bibfield  {author} {\bibinfo {author} {\bibfnamefont {E.}~\bibnamefont {Poisson}}\ and\ \bibinfo {author} {\bibfnamefont {C.~M.}\ \bibnamefont {Will}},\ }\href@noop {} {\emph {\bibinfo {title} {Gravity: Newtonian, Post-Newtonian, Relativistic}}}\ (\bibinfo  {publisher} {Cambridge University Press},\ \bibinfo {year} {2014})\BibitemShut {NoStop}%
\bibitem [{\citenamefont {{Zhao}}\ \emph {et~al.}(2025)\citenamefont {{Zhao}}, \citenamefont {{Tang}},\ and\ \citenamefont {{Xu}}}]{2025EPJC...85...36Z}%
  \BibitemOpen
  \bibfield  {author} {\bibinfo {author} {\bibfnamefont {L.}~\bibnamefont {{Zhao}}}, \bibinfo {author} {\bibfnamefont {M.}~\bibnamefont {{Tang}}}, \ and\ \bibinfo {author} {\bibfnamefont {Z.}~\bibnamefont {{Xu}}},\ }\href {\doibase 10.1140/epjc/s10052-025-13767-0} {\bibfield  {journal} {\bibinfo  {journal} {European Physical Journal C}\ }\textbf {\bibinfo {volume} {85}},\ \bibinfo {eid} {36} (\bibinfo {year} {2025})},\ \Eprint {http://arxiv.org/abs/2411.01979} {arXiv:2411.01979 [gr-qc]} \BibitemShut {NoStop}%
\bibitem [{\citenamefont {{Meng}}\ \emph {et~al.}(2024)\citenamefont {{Meng}}, \citenamefont {{Xu}},\ and\ \citenamefont {{Tang}}}]{2024arXiv241101858M}%
  \BibitemOpen
  \bibfield  {author} {\bibinfo {author} {\bibfnamefont {L.}~\bibnamefont {{Meng}}}, \bibinfo {author} {\bibfnamefont {Z.}~\bibnamefont {{Xu}}}, \ and\ \bibinfo {author} {\bibfnamefont {M.}~\bibnamefont {{Tang}}},\ }\href {\doibase 10.48550/arXiv.2411.01858} {\bibfield  {journal} {\bibinfo  {journal} {arXiv e-prints}\ ,\ \bibinfo {eid} {arXiv:2411.01858}} (\bibinfo {year} {2024})},\ \Eprint {http://arxiv.org/abs/2411.01858} {arXiv:2411.01858 [gr-qc]} \BibitemShut {NoStop}%
\end{thebibliography}%
\end{document}